\documentclass{elsart}	
\usepackage{graphicx}
\usepackage{amsmath}
\usepackage{amssymb}
\usepackage{xr} 

\begin{document}
\date{Accepted in Physica D}

\begin{frontmatter}

\title{Nonlinear dynamics of waves and modulated waves in 1D thermocapillary flows.\\
	II: Convective/absolute transitions.}

\author{Nicolas Garnier\thanksref{atlanta2}\corauthref{nico2}}
\author{Arnaud Chiffaudel}
and
\author{Fran\c{c}ois Daviaud}

\address{Groupe Instabilit\'es et Turbulence,
	Service de Physique de l'Etat Condens\'e, \\
        Direction des Sciences de la Mati\`ere, CEA Saclay, CNRS URA 2464, \\
        B\^at. 772, Orme des Merisiers, 91191 Gif-sur-Yvette, France}

\thanks[atlanta2]{Present address : Center for Nonlinear Science, 
		Georgia Institute of Technology GA 30332-0430.
		{\tt garnier@cns.physics.gatech.edu} }
\corauth[nico2]{Corresponding author}

\begin{abstract}

We present experimental results on hydrothermal waves in long and narrow
1D channels. In a bounded channel, we describe the primary and secondary
instabilities leading to waves and modulated waves in terms of
convective/absolute transitions. Because of on the combined effect of
finite group velocity and of the presence of boundaries, the
wave-patterns are non-uniform in space. We also investigate
non-uniform wave-patterns observed in an annular channel in the presence
of sources and sinks of hydrothermal waves. We connect our observations
with the complex Ginzburg-Landau model equation in the very same way as
in the first part of the paper~\cite{part1}.

\end{abstract}

\begin{keyword}
hydrothermal waves \sep Ginzburg-Landau equation \sep Eckhaus instability \sep 
convective/absolute transition \sep modulated waves
\end{keyword}

\end{frontmatter}
\newpage

\tableofcontents
\newpage

\section*{Introduction}

Non-linear traveling-waves have exhibited a fascinating variety of
behaviors and patterns. Waves systems have been studied in binary-fluid
convection (subcritical traveling waves
bifurcation)~\cite{kol92,baxeat92}, oscillatory instability in low
Prandtl number convection~\cite{janpum92}, oscillatory rotating
convection~\cite{liueck99} and cylinder wake~\cite{lewpro95}. One of the
main source of richness in wave patterns is the existence of two
different regimes: the convective and the absolute
one~\cite{deiss85,coucho99}. This distinction arises when the group
velocity of the waves in non-zero. The convective/absolute transition
between those two regimes leads to critical phenomena of important
relevance in wave-systems. We present here the first complete
description of convective/absolute transitions for traveling waves in
finite a box for both primary and secondary instability onset.

Most 1D physical systems produce right- and left-propagating non-linear
waves \cite{hecsto99}. Non-linear competition and reflections at the
boundaries then lead to a central-source pattern
\cite{crowil89,chiper89,cross88,crokuo92} due to counter-propagating
exponentially growing waves as the first global mode at onset. But the
global mode may also be produced at the convective/absolute transition
as we will illustrate. So far, this phenomenon was described for single
waves~\cite{deiss85,tobpro98,gonern99}, i.e., with broken left-right
symmetry.

Hydrothermal waves \cite{mukchi98,garchi01,garchi02,burmuk01} provide
very interesting and generic systems of traveling waves which can be
modelized by envelope equations such as the complex Ginzburg-Landau
equation (CGL)~\cite{crohoh93,arakra01}. The present paper is the second
part of an article devoted to the connection between hydrothermal waves
and those amplitude equations. We will further refer to the companion
paper~\cite{part1} as I. 

In I, we present the experimental
hydrothermal waves systems, and the periodic solutions obtained in an
annular cell. We will often refer to this paper, but we recall here in
Section~\ref{sec:rec:system} the main characteristics of our systems and
some important results. The present paper is mainly devoted to the case
of a rectangular cell, {\em i.e.}, of a long narrow channel with
non-periodical boundary conditions. We show that hydrothermal waves
realize as in I an ideal supercritical non-linear wave system in
accordance with theoretical predictions in such a bounded geometry:
convective and absolute transitions have to be taken into account to
describe the effect of the group velocity over the onsets of both
primary and secondary instabilities. We also present results in an
annular cell with periodic boundary conditions in the special case where
the periodicity is broken by the pattern itself.

In Section~\ref{sec:rect_seuil1}, we study first the critical behavior at
the primary onset for a finite cell with low-reflection boundaries. When
the wave pattern appear, it is qualitatively very different from the
uniform hydrothermal waves (UHW) pattern observed within periodic
boundary conditions and described in~I, through the governing
equations are the same. This results from the broken Galilean invariance
due to the existence of boundaries. We will show that the first global
mode, instead of being constructed by successive reflections in the
convective regime, results from the onset of absolute instability. The
onset of this mode ---corresponding to the convective/absolute
transition--- is shifted above the value corresponding to convective
instability. This is illustrated in Fig.~\ref{fig:A_vs_DT} of~I and we
will detail qualitatively and quantitatively how the transition occurs.
Some experimental critical exponents are discussed in the framework of
existing theoretical descriptions and a quantitative comparison with
complex Ginzburg-Landau model is proposed.

Section~\ref{sec:rect_seuil2} is devoted to modulated waves. For higher
values of the control parameter, the wave pattern undergoes a secondary
modulational instability; we present the convective/absolute transition
for this instability in the bounded channel. This instability is of the
same nature as the Eckhaus instability occurring on UHW in the annular
channel and described in the companion paper~I: it leads also to a
wave-number selection in the rectangular channel. We will introduce
fronts of spatio-temporal defects and link them to the convective and
absolute nature of the instability.

In section~\ref{sec:sources}, we show that the convective/absolute
transitions have reminiscent effects in the annular geometry, when
sources and sinks exist that break the Galilean invariance as physical
boundaries do. The group velocity term cannot be cancelled out of the
equations anymore where such objects are present. This leads to
qualitative and quantitative behaviors predicted
theoretically~\cite{hecsto99} and observed in our experimental system; we
then confirm the pertinence of the convective/absolute instability
transitions.

\label{sec:rectangle}
\section{The rectangular geometry and the wave model}
\label{sec:rec:system}

The experimental system has been described in section~\ref{sec:setup}
and Fig.~\ref{fig:schema_rect} of the companion paper I~\cite{part1}.
We study the traveling-waves instability of a thermocapillary flow
obtained when applying an horizontal temperature gradient over a thin 
liquid layer with a free surface.
Let us emphasize that we use annular and rectangular channels of both
the same width $10$mm, and that the fluid height is $h=1.7$mm for all
experiments reported in the present paper. The curvature is
negligible~\cite{garnier00}, and we have shown by stability
analysis~\cite{garnor01} that its effects on critical values are of
order $3.10^{-2}$, so both wave pattern reports can be directly
connected. Please note that the channel length will be noted $L_{\rm b}$
for the bounded (rectangular) channel and $L_{\rm p}$ for the periodic
(annular) channel. Without subscript, $L$ will concern the current
channel, and $L^*$ the non-dimensional length $L/\xi_0$ where $\xi_0=5.1$mm
is the coherence length of the pattern (see
section~\ref{sec:model} in I). We have $L_{\rm b}=180$mm and $L_{\rm
p}=503$mm; the aspect ratios in both cells ensure that patterns are
one-dimensional.

The control parameter is the horizontal temperature difference $\Delta T$
between the two long sides of the container. The fluid is observed using 
shadowgraphy and taking care of being in the linear regime (see I),
i.e., we record the shadowgraphic image on a screen located at a distance
much larger than the focal distance of the convection pattern, even for larger values
of the temperature constraint~\cite{garnier00}.
The first bifurcation of the basic thermocapillary flow towards
hydrothermal waves occurs in the annulus for $h=1.7$mm at $\Delta T_{\rm
c} = (3.1 \pm 0.1)$K as described in section~\ref{sec:hydrothermal} of
I. The bifurcated pattern for hydrothermal waves is then a
uniform-amplitude traveling wave of critical wavenumber $k_0 = (0.684
\pm 0.003)$mm$^{-1}$ and critical frequency $f_0 = 0.237$Hz. In the
rectangular geometry, boundaries act as sources or sinks and in fact
both right and left traveling waves are present at threshold. So
hydrothermal waves must be modeled by two slowly varying
amplitudes $A$ and $B$ obeying two complex Ginzburg-Landau
equations:
\begin{eqnarray}
\tau_0 (A_T+s A_X) = \epsilon(1+ic_0) A &+& \xi_0^2 (1+ic_1)A_{XX} \nonumber \\
          \mbox{} &-& g(1+ic_2) |A|^2A - g(\lambda+i\mu)|B|^2A     \nonumber \\
\tau_0 (B_T-s B_X) = \epsilon(1+ic_0) B &+& \xi_0^2 (1+ic_1)B_{XX} \nonumber \\
          \mbox{} &-& g(1+ic_2) |B|^2B - g(\lambda+i\mu)|A|^2B
\label{eq:cgl:rec}
\end{eqnarray}

$\epsilon=(\Delta T - \Delta T_{\rm c})/\Delta T_{\rm c}$ is the reduced
distance from threshold.
Boundary conditions for $A$ and $B$ should be included with this
description. Perturbations are verified to travel at the group velocity
$v_g$ in the rectangular box as well as in the annular one. Please note
that in the CGL Eqs~(\ref{eq:cgl:rec}), $s$ denotes the value of the
group velocity $v_g$ at the convective onset. The coefficients $\tau_0$, $c_0$,
$c_1$, $c_2$, $g$, $\lambda$ and $\mu$ are all real and commented in I.

\begin{figure}
\begin{center}\includegraphics[width=12cm]{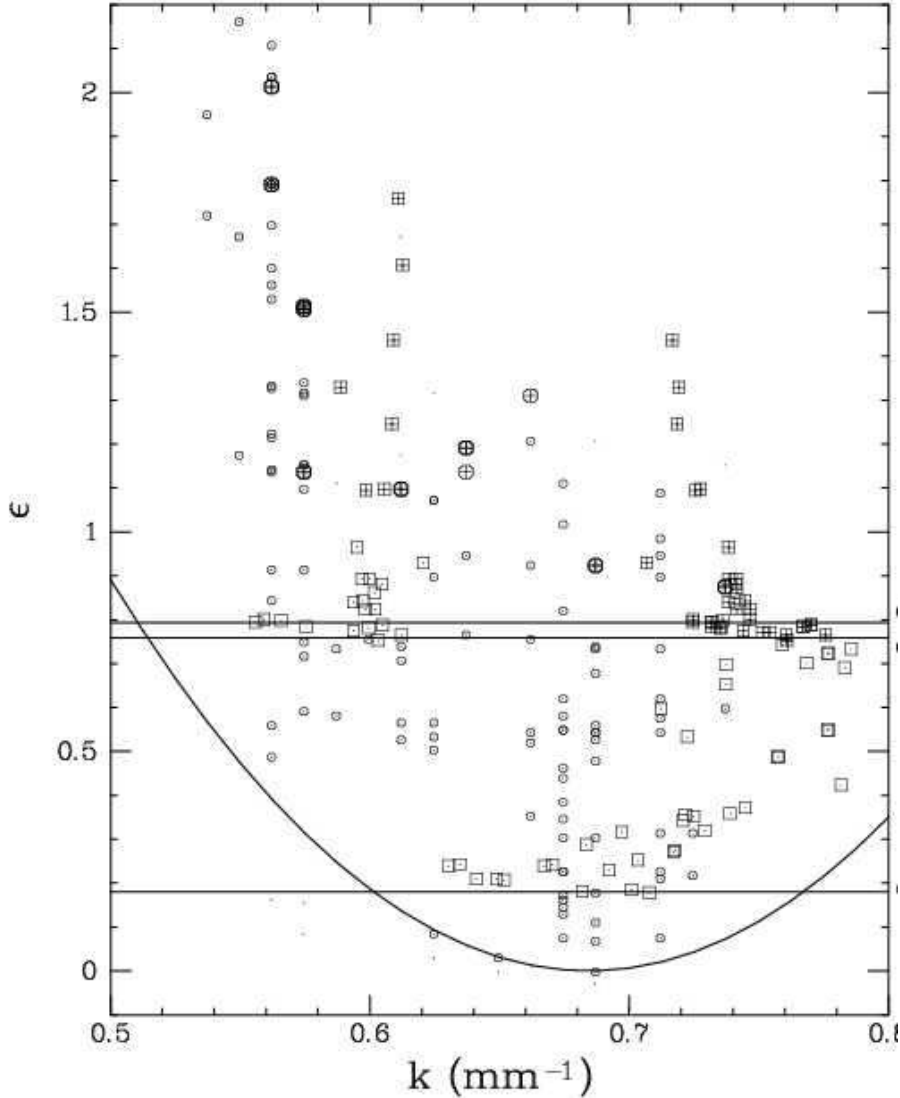}\end{center}

\caption{Experimental stability diagram for hydrothermal waves at
        $h=1.7$mm. Data are presented for both annular ($\circ$) and rectangular
        ($\Box$) geometry. In the annulus, small circles stand for homogeneous
        waves. Modulated waves are depicted with and additional cross $+$. In
        the rectangle, no waves are observed below $\epsilon_{\rm a}=0.18$, 
        and for $\epsilon > \epsilon_{\rm m,c}=0.76$ points are grouped
        on two different vertical branches. Those branches correspond to two 
        different wave trains at the same time within the cell.
        An additional $+$ in the squares denotes that those waves   
        are modulated. $\epsilon_{\rm m,c}$ and $\epsilon_{\rm m,a}$
        stands for the onset of convective and absolute modulational
        instability (Section~\ref{sec:rect_seuil2}).
	The solid parabola is the measured marginal stability curve. }

\label{fig:ballon:ann&rect}
\end{figure}

In order to quantitatively connect results in annular and rectangular
geometry, we propose in Fig.~\ref{fig:ballon:ann&rect} a representation
of stable states in the annulus together with all states obtained in the
rectangle. In this figure, as in all the text in this article,
$\epsilon$ is defined for both experiments using the onset in the
annulus. Wavenumbers are expressed in such a way that they do not depend
on the size of the system, so we use dimensional unit (mm$^{-1}$). In
the annulus, we observe uniform hydrothermal waves (UHW), {\em i.e.},
waves with constant amplitude, wavenumber and frequency and modulated
waves (MW), {\em i.e.}, waves with modulated amplitude, wavenumber and
frequency. The existence of UHW and MW suggest that the wave system may
be described using phase equations. Please note that in the annulus, no
particular selection of the wavenumber exists within the stability zone
of UHW, except the restriction of that the wavenumber must be an integer
when expressed in $(2\pi/L)$ units. In the rectangle, no such
restriction exists and the wavenumber can take any value. 

Let's describe the states observed in the rectangular channel.
First, we note that between $\epsilon=0$ and
$\epsilon=\epsilon_{\rm a}=0.18$, we never observe waves.
This fact will be detailed in
Section~\ref{sec:rect_seuil1} presenting the primary onset.
When increasing the control parameter from the
threshold value $\epsilon_{\rm a}$, the wavenumber $k$ increases
while exploring the whole band of allowed wavenumbers.

Above $\epsilon \simeq 0.45$, a smooth selection process occurs and the wavenumber
$k$ is of order of $21\cdot(2\pi/L_{\rm b}) \simeq 0.73$mm$^{-1}$: this correspond to the
vertical branch on the right of the stability diagram.
For $\epsilon > \epsilon_{\rm m,c}\sim 0.758 \pm 0.01$,
the waves are modulated and two regions are observed in the cell,
corresponding to two different mean wavenumbers. The first region corresponds 
to the previous right branch and the other to the left
branch of the diagram, 
with $k \sim 17\cdot(2\pi/L_{\rm b}) \simeq
0.59$mm$^{-1}$. This selection 
process is rather sharp.
Above $\epsilon_{\rm m,a}$, any state in the cell 
is represented by a point on each of the two branches. 
Between $\epsilon_{\rm m,c}$ and $\epsilon_{\rm m,a}$, 
the left branch is only populated by transients.
The splitting of the
$(k,\epsilon)$ curve in two branches is due to the Eckhaus
instability and will be carefully detailled in section~\ref{sec:rect_seuil2}.

Please note that far from onset, wavenumbers are
either greater or smaller than the critical one and in fact the two
branches observed are away from $k=k_{\rm c}=(0.684 \pm 0.003)$mm$^{-1}$
(Section 2.6 of I). 

Let's also note that in the rectangle, due to the non-periodic 
boundary conditions, the amplitude is forced to vanish or have 
very low values, thus jeopardizing any phase description. Our wave-system 
is such that the boundaries at $x=0,L$ are poorly reflective~\cite{garchi01}:
we have indications that the reflection coefficient is within the range
$10^{-3}-10^{-1}$.

Please note that for sufficiently high $\epsilon$ ($\ge 1$), we observe
the same asymmetrical selection of wavenumbers $k<k_{\rm c}$, as the
annulus. This global selection confirm the relevance of higher order
terms perturbing the amplitude equation (Section~\ref{sec:model} in I).
However, let's emphasize that when considering the primary onset,
{\em i.e.}, small value of $\epsilon$ and small amplitudes $A$ and $B$
of the waves, we may neglect the higher order terms. In contrast,
when presenting the Eckhaus instability leading to the
selection of lower wavenumbers, one should include those higher order
terms to perform a correct analysis of the secondary instability.

The main result from Fig.~\ref{fig:ballon:ann&rect} is the perfect
overlap of the zones of existence. Stable waves ---Stokes solutions of
the CGL eqs (\ref{eq:cgl:rec})--- are observed in the rectangle and in
the annulus in the same region of the $(k,\epsilon)$ plane. Moreover,
modulated waves, and so the Eckhaus instability, occur in the same
regions. Though the spatial extents of the cells are different, they are
both extended and the global wavenumber selection process looks the
same: as $\epsilon$ is increased, the wavenumber is expected to be
lower. The overlap also confirm that coefficients in amplitude equations
are likely to be identical in both geometries (Section \ref{sec:model}
in I). 

\section{Onset of primary wave instability in the rectangle}
\label{sec:rect_seuil1}

We describe here how the wave pattern appears and evolves near the
primary onset. As explained, for small values of $\epsilon$ as those
considered here, higher order terms can be neglected in the
amplitude equations and we believe that usual Complex Ginzburg-Landau
equations for traveling waves are sufficient for the description of the
primary onset. In the following we choose, for clarity, to present the
major wave as $A$ (right-traveling) and the minor wave as $B$
(left-traveling), but the reverse situation has been observed with equal
probability.

\subsection{Description of the patterns}
\label{sec:description}

\begin{figure}
\begin{center}\includegraphics[width=8cm]{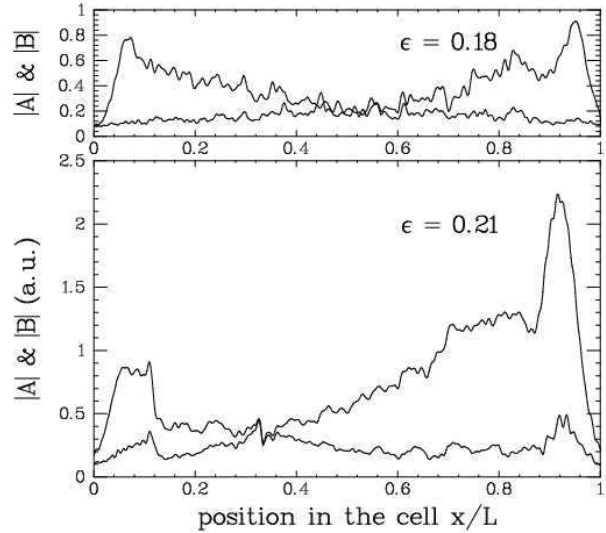}\end{center}
\caption{Amplitude profiles of the traveling waves at different
        $\epsilon$. Top figure presents right ($A$) and left ($B$) amplitude
        profiles for $\epsilon=0.18$ ($\Delta T=3.66$K) and bottom
        figure for $\epsilon=0.21$ ($\Delta T=3.75$K). 
        }
\label{fig:rec:profiles}
\end{figure}

\begin{figure}
\begin{center}\includegraphics[width=14cm]{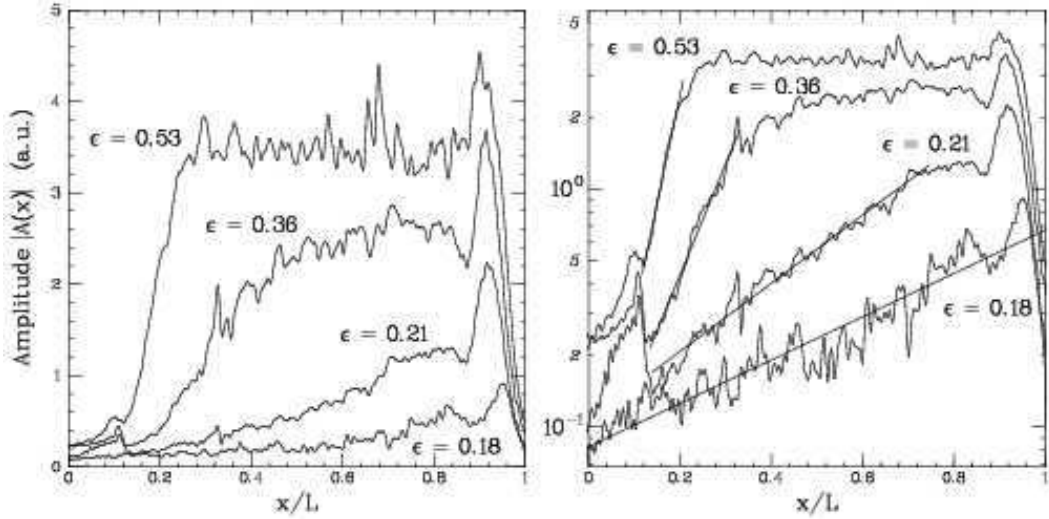}\end{center}
\caption{Amplitude profiles of the dominant traveling wave for
        growing $\epsilon$ up to $0.53$ ($\Delta T=4.75$K). On both
        figures, the data is the same; left: linear scale; 
        right: logarithmic scale. All dominant waves
        are represented as right-traveling waves. Following the wave, we
        encounter three domains: an exponential growth or front, a saturated
        plateau (for larger $\epsilon$), and a sharp wall-mode.
	The linear fits on the lin-log plot give the value of the exponential
	spatial growth rate $\xi_{\rm F}^{-1}$. Similarly, spatial growth $\xi_{\rm
	WM}^{-1}$ and decay rate $\xi_{\rm down}^{-1}$ can be fitted on both sides of the
	wall-mode on the right side of the plot.
        }
\label{fig:rec:profiles_log}
\end{figure}

\begin{figure}
\begin{center} \includegraphics[width=8cm]{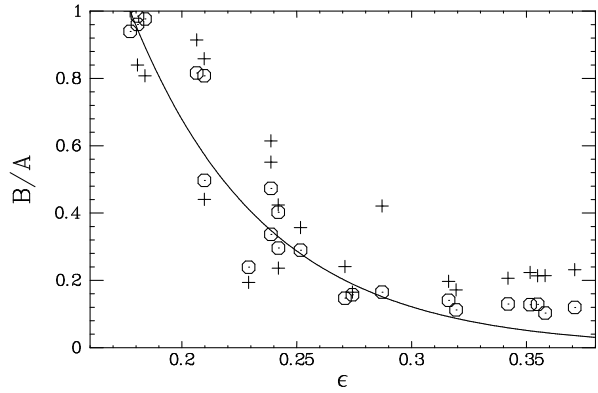} \end{center}
\caption{Amplitude ratio $B/A$ for both averaged amplitude $\langle A \rangle$ 
        over the cell ($\circ$) and maximum amplitude $A_{\rm max}$ ($+$). 
        See text for details. Solid line is an exponential fit.
        As $\epsilon$ is increased higher above onset, the minor 
        wave disappear.}
\label{fig:rec:asymetry}
\end{figure}

Typical amplitude profiles in the rectangle for $A$ and $B$ are shown on
Figs.~\ref{fig:rec:profiles} and~\ref{fig:rec:profiles_log} for various
temperature differences. For $\Delta T< \Delta T_{\rm a}=3.65$K
($\epsilon<\epsilon_{\rm a}=0.177$), no wave is observed and so $A=B=0$.
Just above onset, for $\Delta T = 3.66$K, we observe a symmetric
wave-pattern~(Fig.~\ref{fig:rec:profiles}). As $\Delta T$ is increased 
above $\Delta T_{\rm a}$, {\em i.e.},
$\epsilon$ is increased above $\epsilon_{\rm a}$, one wave becomes
dominant and invade a larger region of the 
cell~(Fig.~\ref{fig:rec:profiles_log}). The waves compete up to
$\epsilon \simeq 0.25$, above which the smallest wave becomes negligible
with respect to the dominant wave~(Fig.~\ref{fig:rec:asymetry}). The
wave envelope of the dominant wave may be seen as composed of three domains:
(i) just after the wall $X=0$ which may also be called source, where
both amplitude $A$ and $B$ are nearly zero, there is a front where the
amplitude is exponentially growing; this is illustrated on
Fig.~\ref{fig:rec:profiles_log}. (ii) after the front, a plateau is
present for higher values of $\Delta T$; (iii) finally, just before the
end wall $X=L$, the amplitude profile has a bump and maximum amplitude
is reached here, in what is called a wall mode.

\begin{figure}
\begin{center}  \includegraphics[width=6.5cm]{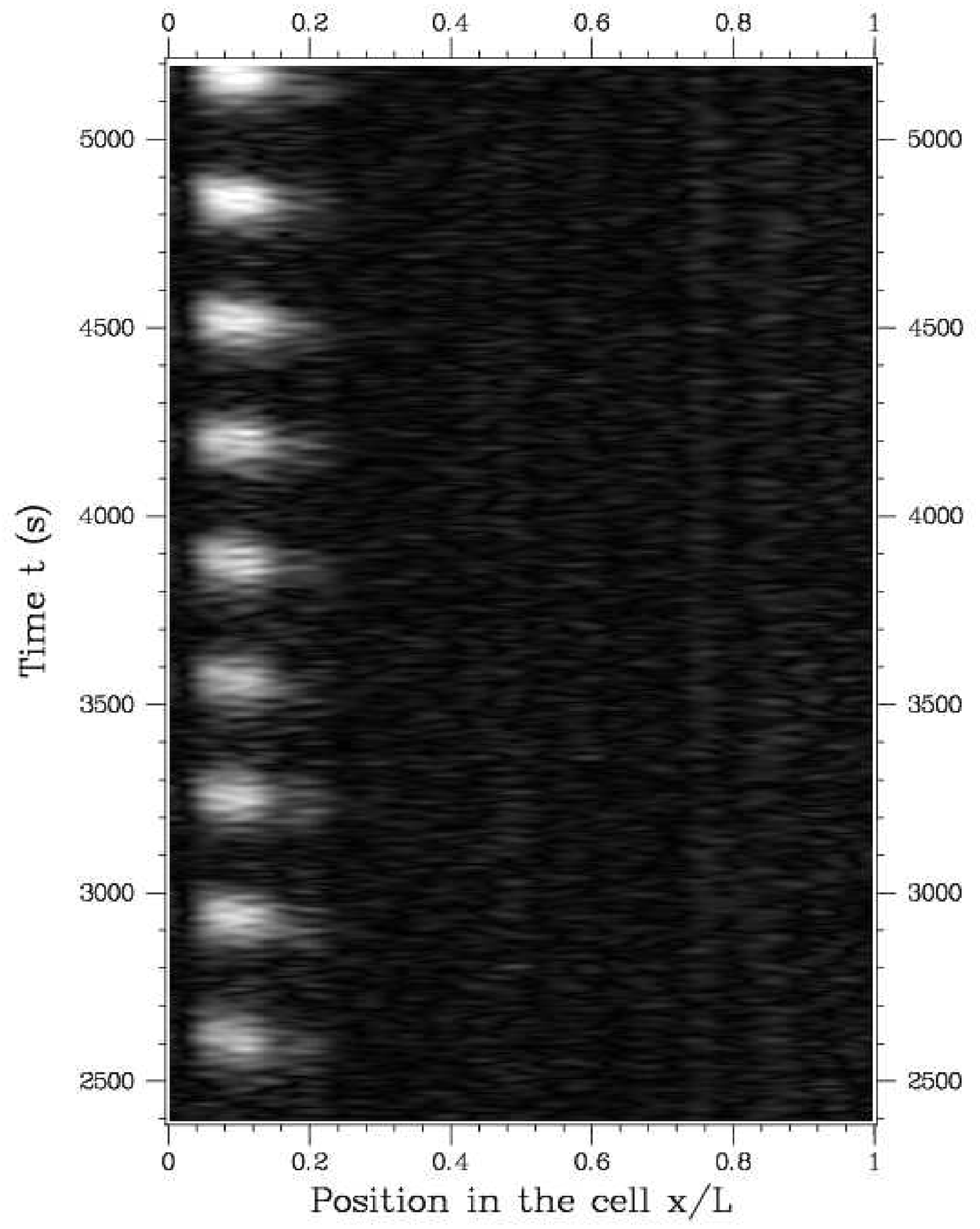}
                \includegraphics[width=6.5cm]{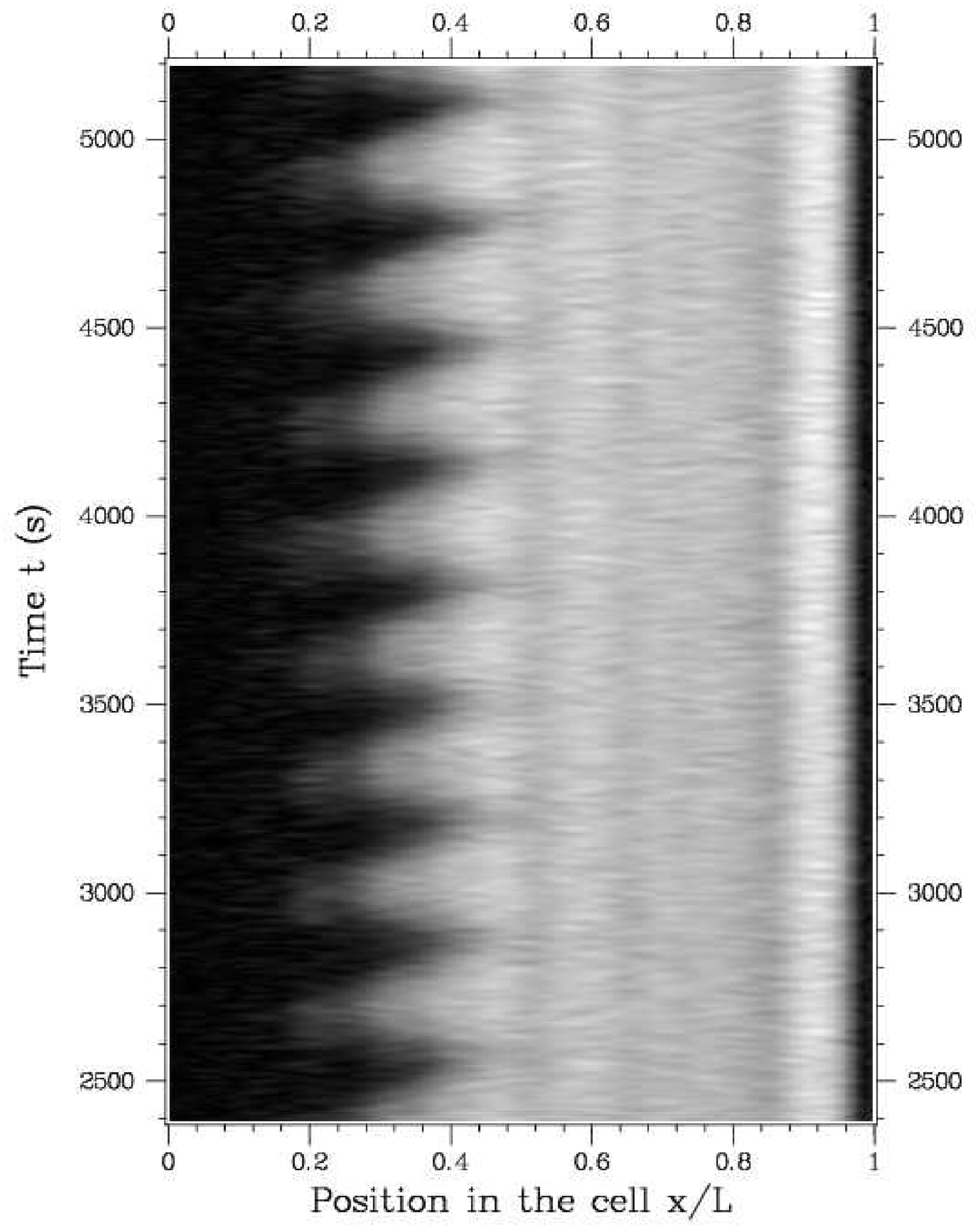}
\end{center}
\caption{Blinking state: spatio-temporal diagram of the local amplitude showing
        the perodical oscillations of the source separating the
        right- and left-traveling waves for $\epsilon=0.36$ ($\Delta T = 4.21$K).
        Left : left-traveling wave. Right : right-traveling wave.
        The wave amplitude is proportional to the gray level of the images
        but for the minor (left) wave, amplitude has been multiplied by
        a factor of 2. Black corresponds to zero amplitude.
        A low-frequency beating is observed with a period around 320s.}
\label{fig:st_blinking}
\end{figure}

In the front region, we can compute for each realization the spatial
growth rate of the amplitude envelope $|A(X)|$ from the source, as shown
in Fig.~\ref{fig:rec:profiles_log}. The front can be
described by an exponential envelope ${\rm e}^{\xi_{\rm F}^{-1}X}$ where
$\xi_{\rm F}^{-1}$ is the growth rate. The characteristic length
$\xi_{\rm F}$ is linked to the front critical behavior~\cite{garchi01}
and will be described in the next paragraph. Please note that the
spatial growth rate in the source region is well defined both close to
the onset, and far from the onset. 

The plateau is vanishing in the vicinity of the threshold and so it
cannot be used to quantify the critical behavior. Quantitative
description of the bifurcation are given in terms of the wall-mode
amplitude and the front spatial growth rate. 

Please note that the amplitude is almost vanishing at the boundaries
$X=0$ and $X=L$. This is an indication for a very low reflection
coefficient in our system~\cite{kapste93}.

When $\epsilon \gtrsim 0.3$, the states are not always stationary and we
observe quasi-periodic realizations corresponding to a beating of minor
and major waves. Those states were called ``blinking states''
\cite{crowil89,finste90,cross88,crokuo92}. Such a realization is
presented in Fig.~\ref{fig:st_blinking}. We believe that the
occurrence of such quasi-periodic patterns is due either to reflections
at the boundaries between both counter-propagating waves and/or to the
nonlinear competition between counter-propagating waves. This scheme is
detailed in Section~\ref{sec:rec:discussion1}.

\begin{figure}
\begin{center}\includegraphics[width=8cm]{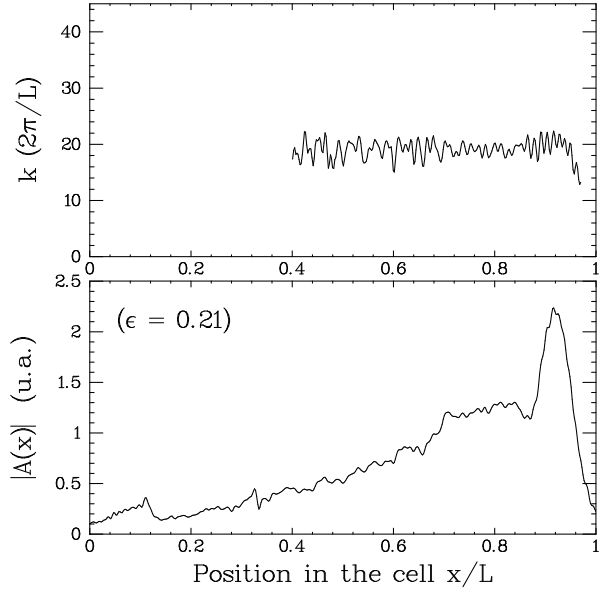}\end{center}
\caption{Spatial profile of the local wavenumber $k$ in the rectangle 
        together with the amplitude profile, for $\Delta T = 3.75$K. 
        Though noisy, the wavenumber is well defined far from the
        source region. 
        $k$ is fairly constant except in the wall mode region 
        where amplitude variations are large.}
\label{fig:rec:profile-k}\end{figure}

\subsection{Quantitative results} 

\begin{figure}
\begin{center}\includegraphics[width=13cm]{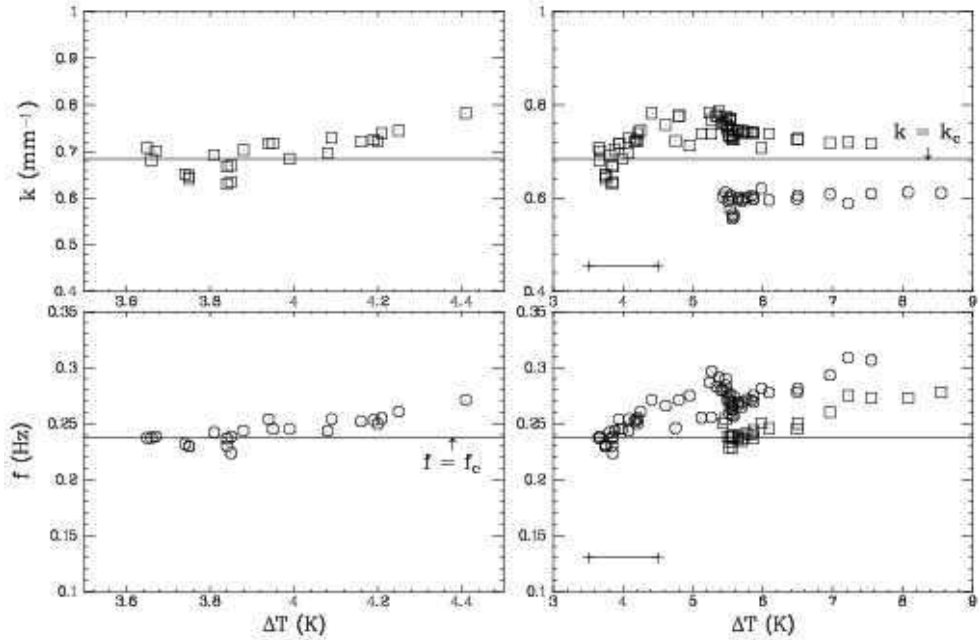}\end{center}
\caption{Frequency and wavenumber in the rectangle {\it vs.} $\Delta T$. 
        Full horizontal lines show the critical values from 
        annular cell data. On the left are shown zooms on 
        the region [3.5K, 4.5K] corresponding to the primary
        instability and represented on the right figures by a segment.
        Different symbols are used for the two branches of solutions 
        for higher values of $\Delta T$; those branches correspond
        to mean wavenumber $21\cdot(2\pi/L_b)=0.73$mm$^{-1}$ and 
        $17\cdot(2\pi/L_b)=0.59$mm$^{-1}$. 
        See also Fig.~\ref{fig:ballon:ann&rect}.
        }
\label{rec_f-k}
\end{figure}

Fig.~\ref{fig:rec:profile-k} presents a typical wavenumber measurement.
We obtain its value everywhere the amplitude is large enough to allow
measurements: we discard regions in which the amplitude is less than
0.5 (a.u.). We then note than except in the wall-mode region where
amplitude variations are large, the local wavenumber is almost constant
in the cell. The mean wavenumber is computed by averaging values far
from the boundaries. It is close to $k \sim 21\cdot(2\pi/L_{\rm b}) \simeq 0.73$mm$^{-1}$.
Measured values of frequency and wavenumber are presented on
Fig.~\ref{rec_f-k}. We note that they match the annulus critical values
at onset, as seen in the stability diagram (Fig.~\ref{fig:ballon:ann&rect}).
For higher values of the
control parameter, frequency and wavenumber are multivalued; this
results from the existence of a new branch of solutions as presented 
on~Fig.~\ref{fig:ballon:ann&rect}. This new branch appears when the primary
wavenumber becomes unstable with respect to modulational
instability (Section~\ref{sec:rect_seuil2}).

\begin{figure}
\begin{center}\includegraphics[width=8cm]{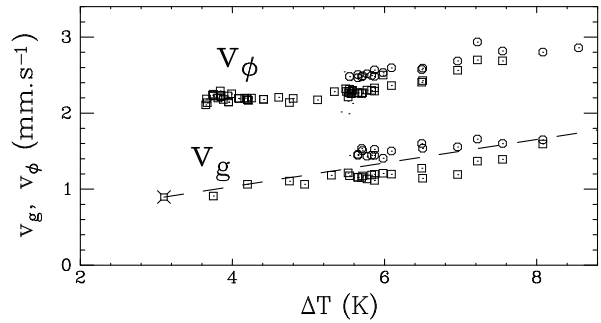}\end{center}
\caption{Phase velocity $v_\phi$ and group velocity $v_g$ in the rectangle
        {\em vs.} $\Delta T$. 
        $\Box$ stands for waves of mean wavenumber close to $21\cdot(2\pi/L)$
        (the right branch in Fig.~\ref{fig:ballon:ann&rect})
        and $\circ$ for  waves of mean wavenumber close to $17\cdot(2\pi/L)$
        (the left branch in Fig.~\ref{fig:ballon:ann&rect}).
        The group velocity is always finite and lower than the phase velocity.
        The symbol at $\epsilon=0$ stands for $s$, the group velocity
        at onset in periodical geometry and the dashed line recalls the
        linear fit obtained in periodical geometry (Fig.~\ref{fig:vg_annulus} in I).}

\label{fig:rec:velocity}\end{figure}

Computing the velocity at which perturbations are advected allows us to
compute the group velocity. The corresponding data is
represented in Fig.~\ref{fig:rec:velocity}, togheter with the phase
velocity of the waves. Here again, for clarity, we distinguish between
waves of different wavenumbers, {\em i.e.}, we distinguish between the two
branches in the stability diagram. Of importance is the fact that the
group velocity is always finite and large. Note that the value of
$v_g$ at $\epsilon_{\rm a}$ is the same as the one at onset in the
annulus, {\em i.e.}, $v_g(\epsilon_{\rm a})=s$. The phase velocity is about
twice the group velocity.

\begin{figure}
\begin{center}\includegraphics[width=13cm]{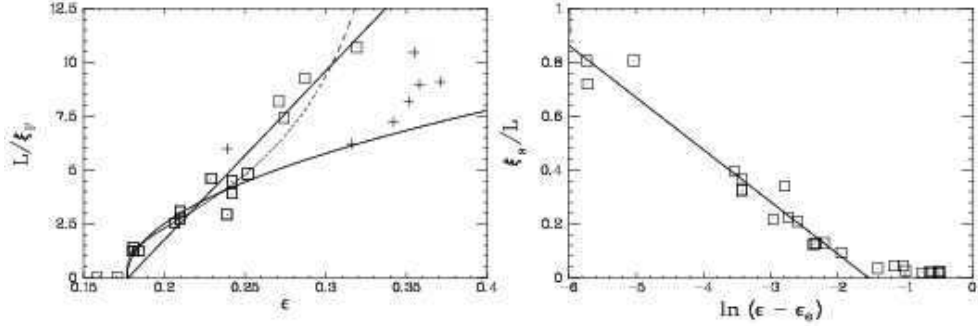}\end{center}
\caption{Critical behavior of the spatial growth rate characterizing the
        front region. Left graph shows experimental data ($\Box$ for
        stationary states and $+$ for blinking states) togheter
        with a linear fit, a square root fit (solid lines) and a
        logarithmic fit (dashed line). Fits are performed on stationary
        data ($\Box$) only. The graph on the right present
        the logarithmic fit
        $\xi_{\rm F} \propto \ln(\epsilon - \epsilon_{\rm a})$. See text for details.}
\label{fig:rec:front}
\end{figure}

Let's now present the critical behavior of the front using the spatial
growth rate $\xi_{\rm F}^{-1}$ defined in the previous paragraph. On
Fig.~\ref{fig:rec:front} are presented such data and possible fits.
Following the simplest intuition, we can propose a linear fit for the
data; this leads to an estimation for the threshold at 3.65K. A closer
observation of the experimental points at 3.66K suggests this may not be
the best description very close to onset: we have conducted several
experiment, at $\epsilon=\epsilon_{\rm a}$, and all led to a small but
finite value of $\xi_{\rm }$, larger than the error bar. We also tried a
square-root law, pertinent close to the onset but jeopardized for values
of $\epsilon$ a bit larger. Following the authors of
Refs~\cite{chocou99,gonern99}, we then searched for a linear relation
between $\xi$ and $\ln(\epsilon-\epsilon_{\rm a})$ and found a better
agreement for all experimental points within a broader range, and most
of all close to the threshold. The physical meaning of this logarithmic
critical behavior will be discussed below.

\begin{figure}
\begin{center}  \includegraphics[width=14cm]{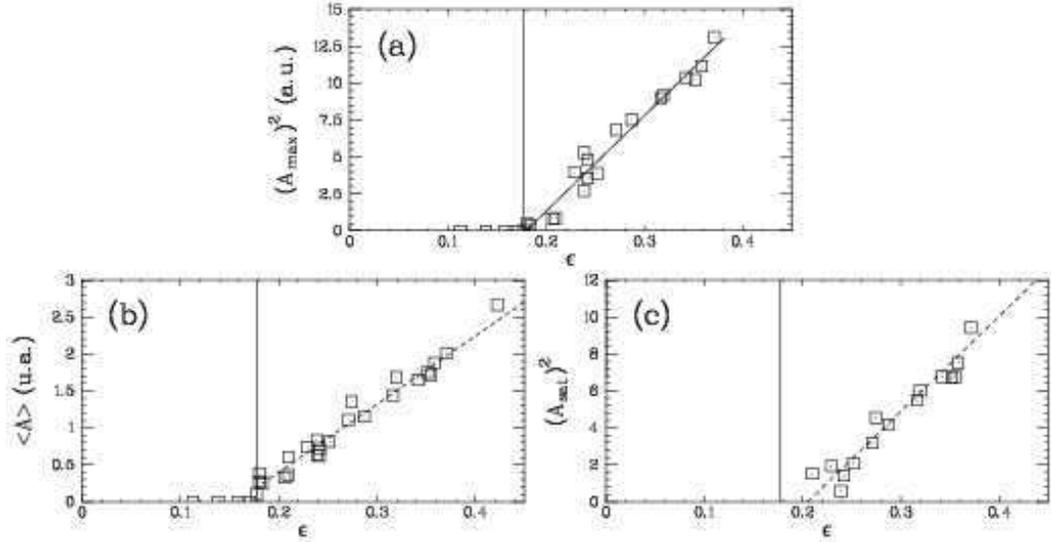} \end{center}
\caption{Critical behaviors in the rectangle for
        (a): the maximum amplitude $A_{\rm max}$, {\em i.e.},
	     the modulus of the wall-mode,
        (b): the mean amplitude $\langle |A(X)| \rangle_{[0,L]}$, and
        (c): the amplitude of the plateau.
        The vertical line represent the onset value $\epsilon_{\rm a}=0.18$ 
        ($\Delta T=3.65$K).
        Only the amplitude of the wall-mode has a critical
        behavior in accordance with the observed onset value.  }
\label{fig:rec:critic}
\end{figure}

Let's look at the amplitude in Fig.~\ref{fig:rec:profiles_log} and
search to define another order parameter of the transition. The
amplitude being non uniform, two quantities are easily extracted: the
averaged and the maximum value of the amplitude profile, both
represented in Fig.~\ref{fig:rec:critic}. The average amplitude $\langle
|A(X)| \rangle_{[0,L]}$ evolves linearly with respect to $\epsilon$
(Fig.~\ref{fig:rec:critic}b), but shows a small finite step at
$\epsilon_{\rm a}$. On the other hand, the maximum $A_{\rm{max}}$ which
occurs near the downstream boundary, at the top of the wall-mode,
behaves like $(\epsilon- \epsilon_{\rm a})^{1/2}$
(Fig.~\ref{fig:rec:critic}a). We believe it to be the order parameter of
this supercritical bifurcation. Please note that this quantity is
studied here very carefully around its onset $\epsilon_{\rm a}$, while a global
view is given in I over a wide range of $\epsilon$ in order to compare
to the annular geometry which bifurcate at $\epsilon=0$.
Finally, as noted in the previous
paragraph, the saturation amplitude within the plateau region cannot be
measured close to the transition. This does not allow one to conclude
anything about the threshold value; a linear fit of the squared
amplitude versus $\epsilon$ even leads to an incorrect value of the
onset (Fig.~\ref{fig:rec:critic}c).

Last, the ratio of the two amplitudes $A$ and $B$, plotted on
Fig.~\ref{fig:rec:asymetry}, behaves like $\exp(-\alpha
\frac{\epsilon-\epsilon_{\rm a}}{\epsilon_{\rm a}})$ with $\alpha=3.0$.
This is observed both for the averaged amplitudes $(\langle |A| \rangle,
\langle |B| \rangle)$ and the maximum (wall-mode) amplitude $(A_{\rm
max},B_{\rm max})$. The decrease of the minor wave is effective as soon
as $\epsilon > \epsilon_{\rm a}$; this is different from the behavior
observed in~\cite{crowil89}. It denotes a strong competition between the
two waves. Due to the power law increase of the maximum amplitude, this
reveals the exponential disappearance of the minor wave when $\Delta T$ is
increased.

\subsection{Discussion}
\label{sec:rec:discussion1}

\subsubsection{Onset shift}

The onset shift between the annular cell and rectangular cell
experiments is well interpreted using the convective/absolute
transition. This shift along with the critical behavior of the front
have been described in~\cite{garchi01,garchi00}. On
the hydrodynamic point of view, a linear stability analysis of the
thermocapillary problem in a finite box~\cite{priger97} has revealed a
particular spatial growth rate ---an envelope of the main unstable
mode--- at onset, together with a shift of this onset from the value
computed in infinite geometry.

The shift of the onset is very large: $0.18$ in $\epsilon$ or $0.55$K in
$\Delta T$. Usual finite size effect are known to be of order
$(\pi/L^*)^2$, i.e., $0.01$ in $\epsilon$ for the $L^*=35$ rectangular
channel. This law has been verified with a good accuracy for
hydrothermal waves in a variable rectangular cell~\cite{javier99}.
Another difference between the two cells is the curvature. This can be
quantified by a linear stability analysis of the thermocapillary problem
\cite{garnier00,garnor01}: in the annulus, the curvature increases the
onset by $0.03$ in $\epsilon$. None of these effects explains the $0.18$
shift.

So what in the rectangle makes the first global mode, or self-excited
wave, observed above $\epsilon_{\rm a}=0.18$ ? Below $\epsilon_{\rm a}$,
no waves are observed and we unsuccessfully tried to trigger wave-trains
with mechanical perturbations (but not with thermal perturbations which
should be more efficient as tested within a hot wire
experiment~\cite{vindub97}).

\subsubsection{Global eigenmode of the CGL model}

In the periodic channel, once the convective onset is crossed, a
traveling wave (TW) self-organizes after successive rounds in the cell:
the basic uniform TW is a global mode and both convective and absolute
onset collapse~\cite{deiss85}. This corresponds mathematically to the
Galilean invariance of the problem in the annulus : one can eliminate
the group velocity $s$ by studying the problem in a frame moving at
velocity $s$. In the non-periodic channel however, $s$ is finite and the
problem has to be solved in the laboratory frame. The first global mode
is then observed when the growth rate is large enough not to have the
waves envelope advected away by the group velocity. This correspond to
the transition between convective and absolute instabilities of the
primary wave-pattern. This transition occurs in the complex Ginzburg-Landau
equation when $\epsilon$ reaches the critical value $\epsilon_{\rm abs}$ given by:

\begin{equation}
\epsilon_{\rm abs} = \frac{1}{1+c_1^2} \left( \frac{s \tau_0}{2 \xi_0} \right)^2
\label{eq:epsilon_abs}
\end{equation}

Below $\epsilon_{\rm abs}$, waves are convectively unstable in infinite
geometries~\cite{huemon90,Sandstede:2000} and no global mode exists in
finite non-periodical geometry~\cite{tobpro98}. For
$\epsilon=\epsilon_{\rm a}$, the global mode in a bounded system of size
$L$ is of the form:

$$ A(x) = A_{\rm gm} \sin\left(\pi\frac{X}{L}\right) e^{(1-ic_1)\xi_{\rm abs}^{-1}X}
\qquad~\xi_{\rm abs}=(1+c_1^2)\frac{2\xi_0^2}{s\tau_0}
$$

\noindent In the vicinity of this threshold, the global mode amplitude is
predicted to behave as $A_{\rm gm} \propto (\epsilon-\epsilon_{\rm
a})^{1/2}$.

We proposed~\cite{garchi01} that our system exhibit the transition from
convective to absolute instability at the onset of the waves in the
rectangular cell. This conjecture can be summarized as:

\begin{equation}
\epsilon_{\rm a} = \epsilon_{\rm abs}.
\label{eq:rec:conjecture}
\end{equation}

\begin{figure}
\begin{center} \includegraphics[width=13cm]{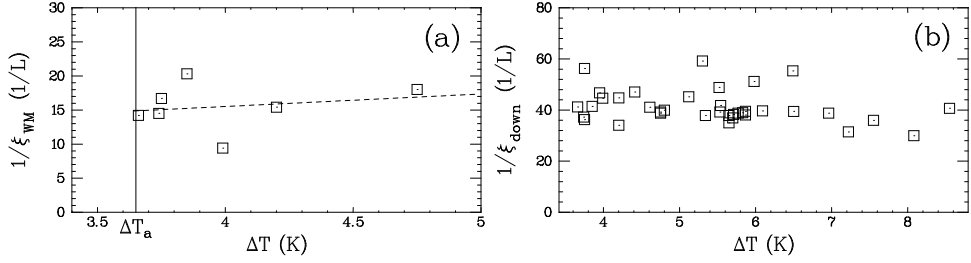} \end{center}
\caption{(a): spatial growth rate $\xi_{\rm WM}$ of the wall-mode
        measured upstream to the wall-mode as the spatial growth rate of the bump.
        (b): spatial damping rate $\xi_{\rm down}$ of the envelope of the wave
        measured downstream of the bump, close to $X=L$.
        $\xi_{\rm WM}$ and $\xi_{\rm down}$ are almost constant with respect
        to the control parameters.}
\label{fig:rec:xi_wm}
\end{figure}

The spatial structure of the predicted global eigenmode is an
exponential growth with a spatial growth rate $\xi_{\rm abs}^{-1}$
independent of $\epsilon$. The main feature of the observed pattern,
however, is the fast varying $\xi_{\rm F}$ of the front region
(Fig.~\ref{fig:rec:front}). Very close to $\epsilon_{\rm a}$, we observe
that the wall-mode is well visible (Fig.~\ref{fig:rec:profiles_log})
This wall-mode shows a spatial growth, measured between the plateau and
the maximum, which is independent of $\epsilon$
(Fig.~\ref{fig:rec:xi_wm}a). This measure is difficult to perform
because the wall-mode growth is almost hidden by the front mode, but it
appears clearly that the wall mode is a good candidate for the global
eigenmode. Quantitative measurements \cite{garchi01} confirms this
hypothesis allowing to measure the value of characteristic time
coefficient $\tau_0$.

After the maximum of the wall-mode, one can finally measure a spatial
damping rate $\xi_{\rm down}^{-1}=(39\pm4)L^{-1}=(1.36\pm0.14)$mm$^{-1}$ for
the envelope of the wave in the wave sink in the vicinity of $X=L$
(Fig.~\ref{fig:rec:xi_wm}b). Within the error bars, this quantity is
independent of $\epsilon$, and is comparable to the wavelength of the
hydrothermal wave. This results is similar to the observation of the damping
in the core of wave sinks by Pastur {\it et al.}\cite{paswes01}. One may
suspect non-adiabatic effects to be responsible for this property. CGL
may not be the right model to describe the sink cores.

Note also that for higher values of $\epsilon$, the
theory~\cite{tobpro98} predicts a wavenumber selection by the front,
exactly as we observe: $k \simeq 21(2\pi/L) \simeq 0.73$mm$^{-1}$
instead of $k_c=0.68$mm$^{-1}$.

\subsubsection{Effect of wave reflections}

This scenario is not the only way to explain a shift in the
threshold. Cross et al.~\cite{cross86,cross88,crokuo92} proposed a
different mechanism involving the two opposite traveling waves and their
mutual reflections at the boundaries $x=0$ and $x=L$. It is found that
waves can be observed only for $\epsilon > \epsilon_{\rm r}$ where:

\begin{equation}
\epsilon_{\rm r} = -s \tau_0 L^{-1} \ln(r) + {\rm O}(L^{-2})
\label{eps_r}
\end{equation}

where $0 \le r \le 1$ is a real number, interpreted as a reflection
coefficient of one wave into the opposite one at the boundary. Martel and
Vega~\cite{Martel:96,Martel:98} explored this theory much further,
explaining how a global mode is constructed for two waves in a bounded
geometry. Their analysis also used the reflection coefficient $r$
between the two waves at the boundaries and the nonlinear coupling
between the two waves. They recovered the experimental succession of
states described by Croquette and Williams~\cite{crowil89}, {\em i.e.},
the existence of a full range in $\epsilon$ where the two-waves pattern
is symmetrical. Furthermore, they described the secondary instability;
but they discarded the effect of the group velocity, {\em i.e.}, the
convective/absolute transition, through it is of great importance in
finite boxes~\cite{tobpro98}. 

The conjecture~(\ref{eq:rec:conjecture}) is equivalent to $\epsilon_{\rm
abs} < \epsilon_{\rm r}$ for the hydrothermal waves system. Owing to
poor reflections in the rectangular cell with Plexiglas ends, we believe
that this is pertinent and that $\epsilon_{\rm r}$ is large. 

Finally, another experimental fact confirms our conjecture. The
experimental sequence of competition between right- and left-wave with
$\epsilon$ (Fig.~\ref{fig:rec:profiles}, \ref{fig:rec:profiles_log}
and \ref{fig:st_blinking}) is the following: symmetric
state $\rightarrow$ asymmetric state $\rightarrow$ quasi-periodic
blinking state $\rightarrow$ ``filling'' state, i.e., a single wave.
This sequence fits the description in case of reflection-controlled
mechanism \cite{cross88,crokuo92,Martel:96,Martel:98} except that the
initial symmetrical pattern exists only for $\epsilon = \epsilon_{\rm
a}$ instead of existing over a finite range as is the case in the
reflection-controlled mechanism, and observed experimentally in that
case~\cite{crowil89,crokuo92}.

Blinking states are observed in our experiment for
$\epsilon \gtrsim 0.3$, we may suggest that this value could be an upper
estimate for $\epsilon_{\rm r}$. It can also be interpreted as the fact
that nonlinear interactions become important for such higher values of
the control parameter.

At the present time, a complete model taking into account both the
convective/absolute transition and the role of the reflection
coefficient and the nonlinear coupling remains to be done in the case of
a two-waves system.

\subsubsection{The nature of the convective/absolute transition}

The last question we wish to address concerns the nature of the
convective/absolute. We observed that the front size behaves
logarithmically around $\epsilon_{\rm a}$ (Fig.~\ref{fig:rec:front}).
This behavior has been detected experimentally by Gondret {\it et al.}
\cite{gonern99} in nonlinear surfave waves and is the signature, in the
sense of Chomaz and Couairon \cite{chocou99}, of a nonlinear
convective/absolute transition. A linear transition
\cite{tobpro98,chocou99} would show the front size to behave as
$(\epsilon-\epsilon_{\rm a})^{-1/2}$. This has a practical consequence on
the experimental observation: since all derivatives of the spatial
growth rate are infinite at $\epsilon_{\rm a}$, the front appears much
faster with $\epsilon$ than the eigenmode. This explains why the
eigenmode is almost hidden by the front and may be detected only as a
faint wall-mode in the close vicinity of $\epsilon_{\rm a}$.

For a better observation of the eigenmode one can suggest to increase
the length $L$ of the channel. The results of Pastur {\it et al.}
\cite{paswes01} for a long hot-wire experiment could have lead to such
observation. However, since $\epsilon_{\rm r}$ decreases with $L$
(Eq.~\ref{eps_r}), these authors report another result: below
$\epsilon_{\rm a}$ (refered to by $\epsilon_{\rm so}$), they observe
fluctuating sources due to the amplification of experimental noise by
the convective instability. Near $\epsilon_{\rm a}$, a crossover leads
to non-fluctuating sources, comparable to the front we observe: the
sources are located in the middle of the channel and not at one end, but
their size also critically depend of $\epsilon$. However, since waves
and sources are already present in the convective region below
$\epsilon_{\rm a}$, the transition itself is also hidden and its nature
cannot be detected.

By increasing the length of our cell and/or by reducing the length of
Pastur {\it et al.} cell \cite{paswes01} we are confident that one could
make the connection between the two experimental observations: to
observe the eigenmode as clearly as we do below a certain critical channel
length, and then analyse how the convective wave becomes visible below
$\epsilon_{\rm a}$ when $L$ gets long enough for $\epsilon_{\rm r}$ to
become smaller than $\epsilon_{\rm a}$.

\section{Onset of secondary modulational instability in the rectangle}
\label{sec:rect_seuil2}

Experiments on non-linear traveling waves have been frequently carried
in annular cells \cite{janpum92,kol92,lewpro95,liueck99,mukchi98} for
the simplicity of the underlying wave pattern. In such periodic
geometries, despite an eventual shift on the
onset~\cite{tucbar90,tucbar91}, the Eckhaus instability is always
absolute. Nevertheless, this instability, as the primary one, may be
convective when the group velocity cannot be canceled out of the
equations~\cite{Weber:92,Aranson:92}. Moreover, the main specificity of
our wave-system is to become Eckhaus unstable for increasing values of
the control parameter, {\em i.e.}, as a first step on the route to
spatio-temporal chaos \cite{part1,mukchi98}. 

As shown in Fig.~\ref{fig:ballon:ann&rect}, this secondary instability
is non-symetrical in $k-k_c$; we proposed in Section~\ref{sec:model} of I that
higher order terms should be included in the amplitude equation. Such
terms are of importance when considering secondary instability because
$\epsilon$ is then no longer small.

Evaporation is limiting the duration of experiments and refills strongly
perturb the patterns. So, in order to get long data series close to
onsets, we generally used a protocol in which refills are made just
before acquisition starts and long thermal stabilization time are
avoided: the temperature gradient is first established in the vessel,
the cell is refilled and the fluid is agitated to break the thermal
gradients. Within a few seconds the hydrothermal waves reappear without
history, {\em i.e.}, without any special values of the wavenumber, or
any special position of dislocations within the cell. However, such a protocol
makes it difficult to test the presence of hysteresis at the transitions
with control parameter $\epsilon$. 

\subsection{Wave system}

From the previous section we know that for $\epsilon > 0.45$ ($\Delta T
> 4.5$K), a single wave is present in the cell with constant
amplitude, wavenumber and frequency (Fig.~\ref{fig:rec:profiles_log}).
This pattern constitutes now our basic state and the present section
focus on the secondary instability of this single wave train for
higher values of $\epsilon$. We choose it to be a right-traveling wave
and write its amplitude as $A$.

In the absence of modulations, this wave train is uniform. This is
illustrated in Fig.~\ref{fig:conv-abs}a: the local and instantaneous 
wavenumber is homogeneous in the cell and constant in time. When modulations
are present, as in Fig.~\ref{fig:conv-abs}b, we perform a second demodulation
to compute the amplitude of this modulation, as well as its 
wavenumber and frequency. This second Hilbert transform is computed
over a spatio-temporal image of the carrier phase derivative (wavenumber
or frequency).

As the group velocity is finite, all perturbations, including the
modulational ones, are advected. We will show the relevance of a new
object, namely a front: a dislocations front or a modulations front. In
periodic conditions (section~\ref{sec:annulus} of I), this modulational
instability occurs at the lowest possible wavenumber $K_{\rm
mod}=2\pi/L_{\rm p}$~\cite{mukchi98,tucbar90}: it is strictly an
Eckhaus~\cite{Eck:65} instability. In the rectangular channel, we will
also refer to Eckhaus instability, although the wavenumber of the
modulational instability modes are somewhat bigger: typically $K_{\rm
mod} \sim 4 \! \cdot \! (2\pi/L_{\rm b})$; it is a finite wavenumber
instability.

In the following, we describe the observed regimes starting from the
absolute one obtained for higher values of the control parameter; we
then present the convective and seemingly stable regimes for smaller
values of $\epsilon$.

\subsection{Absolute instability and corresponding states}

\begin{figure}
\begin{center}  \includegraphics[width=8cm]{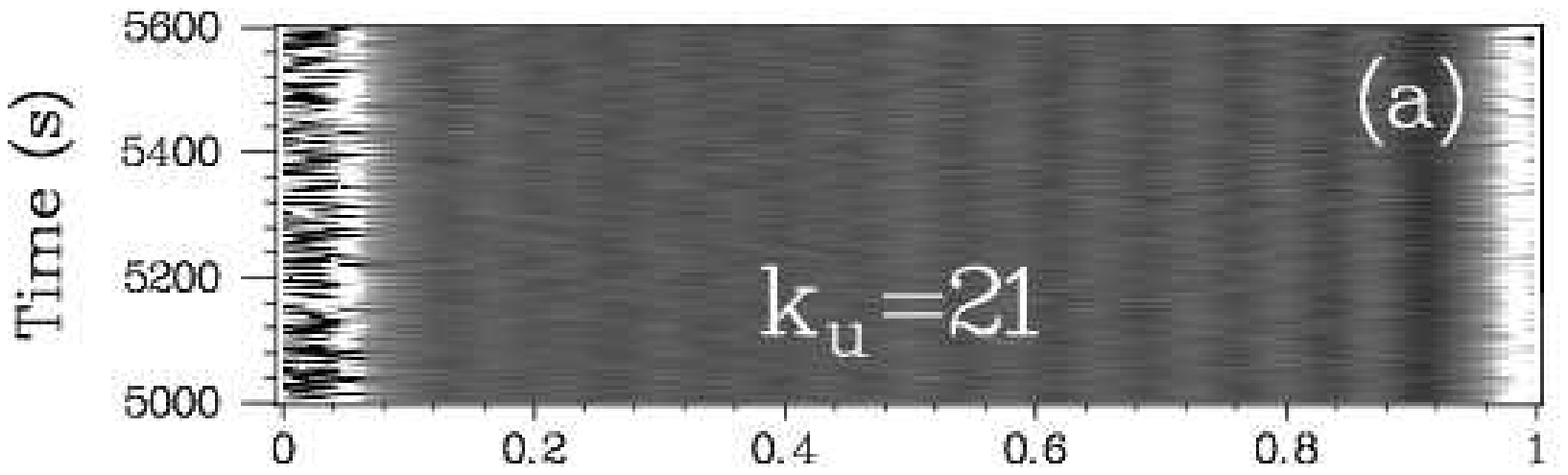}
                \includegraphics[width=8cm]{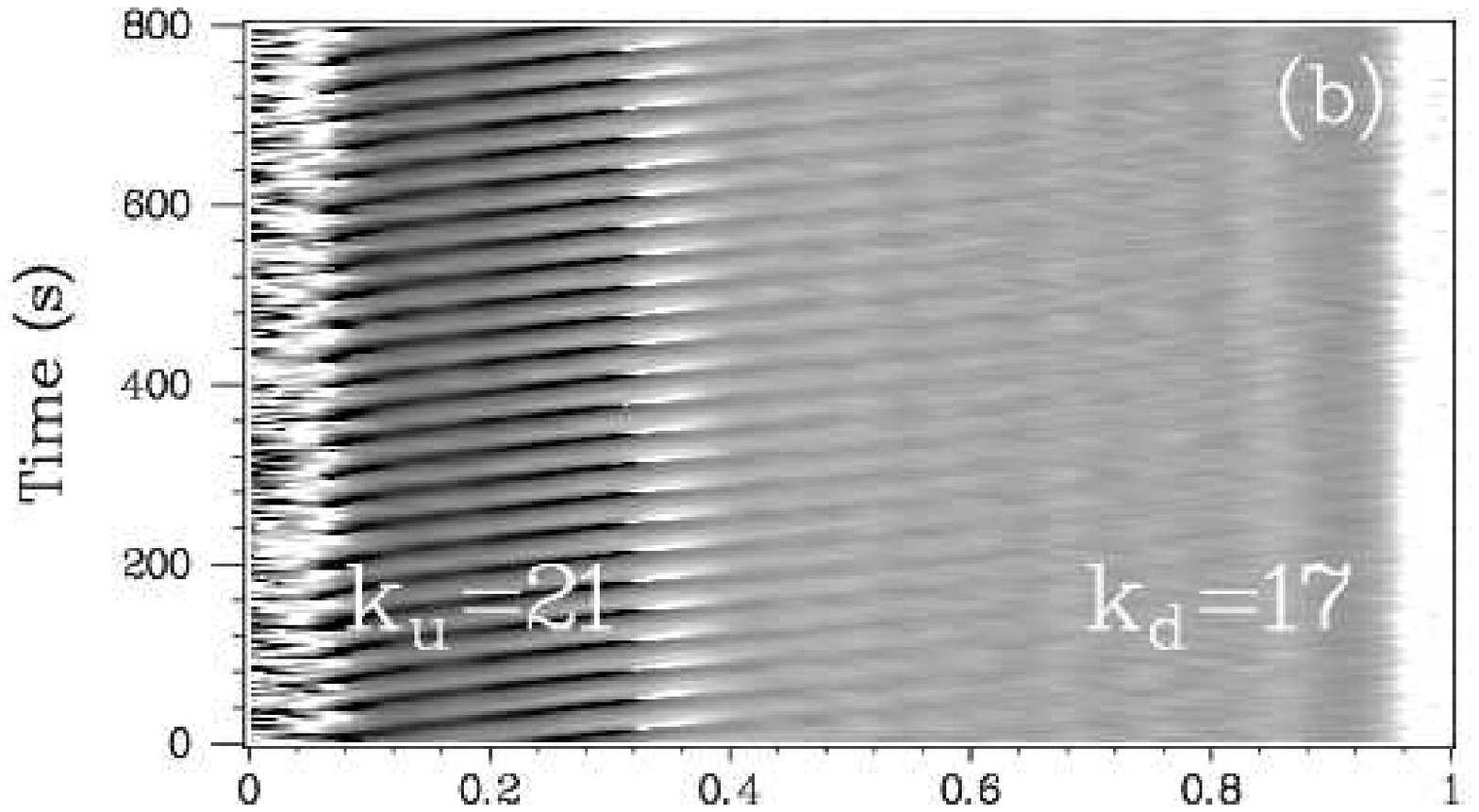}
                \includegraphics[width=8cm]{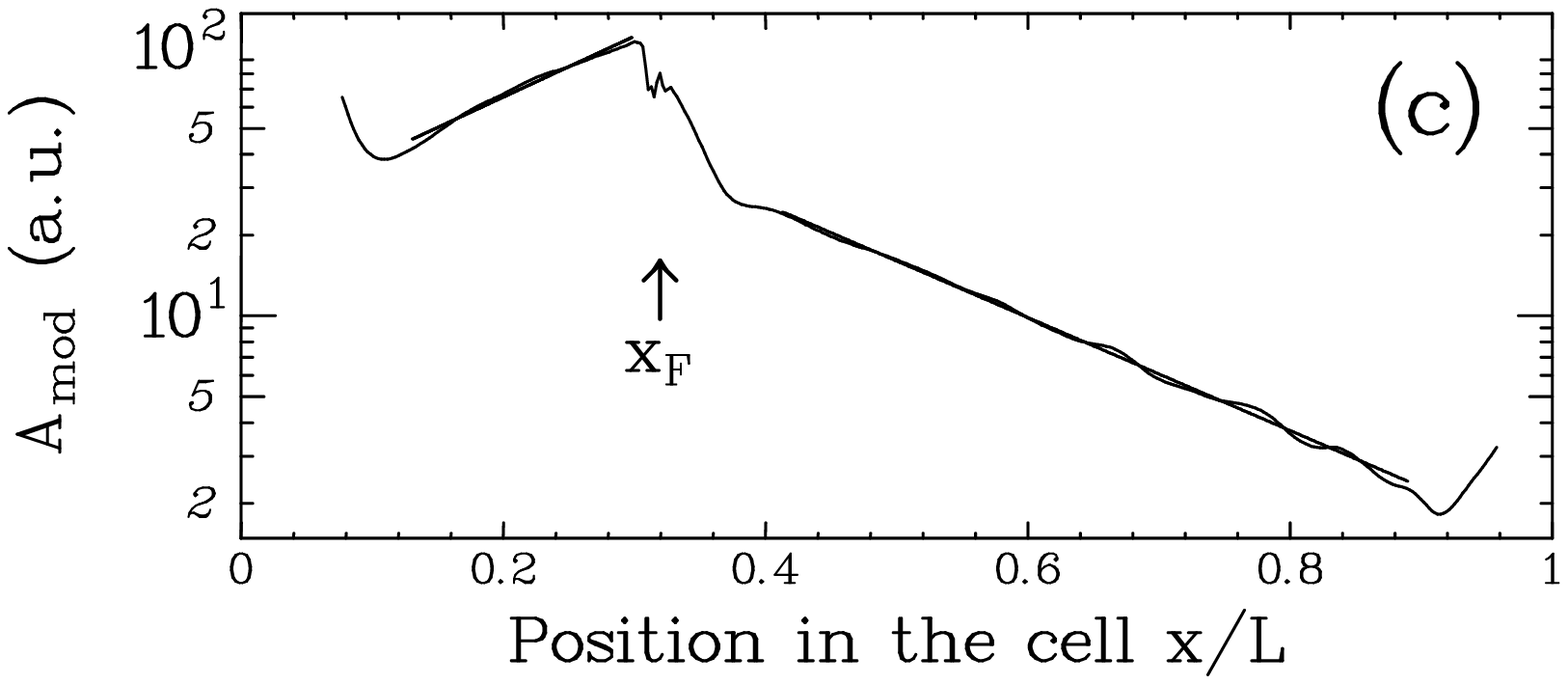}
\end{center}

\caption{Spatio-temporal diagrams of the local and instantaneous
wavenumber $k(x,t)$ of the wave: temporally stabilized regimes for (a)
$\epsilon=0.79$, $\Delta T=5.54$K and (b) $\epsilon=0.82$, $\Delta
T=5.65$K. The waves propagate from left to right. The mean wavenumber
can be estimated visually by the mean gray level and is labeled (units
$2\pi/L$) in the upstream ($k_{\rm u}\simeq 21\cdot(2\pi/L)$) and
downstream ($k_{\rm d}\simeq 17\cdot(2\pi/L)$) regions. A uniform
wavenumber (a) corresponds to an homogeneous state and illustrates both
stable and convective regimes below $\Delta T_{\rm m,a}$. It is the
asymptotic regime obtained after the transient shown in
Fig.~\ref{fig:conv}, left. The modulated state (b) is the global mode of
the Eckhaus instability. Each black to white transition of the
wavenumber value at $x_F/L=0.32$ is due to a phase jump in the core of a
defect. The defect front is stable along time. (c) By Hilbert
demodulation of phase-gradient image (b) we get the spatial profile of
the amplitude $A_{\rm mod}$ of the modulation, presented here in
logarithmic units.}

\label{fig:conv-abs}
\end{figure}

\begin{figure}
\begin{center}  \includegraphics[width=6.5cm]{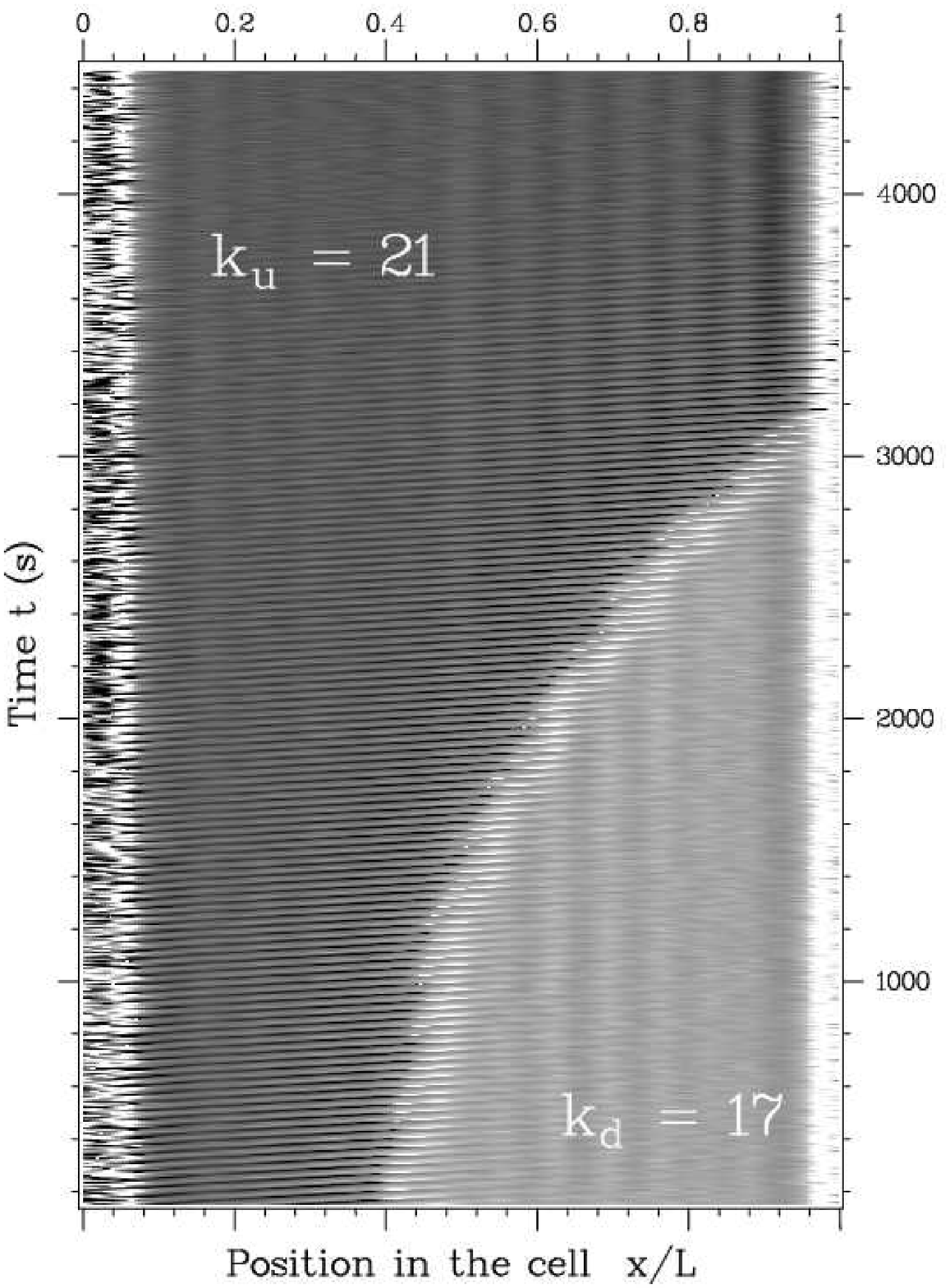}
                \includegraphics[width=6.5cm]{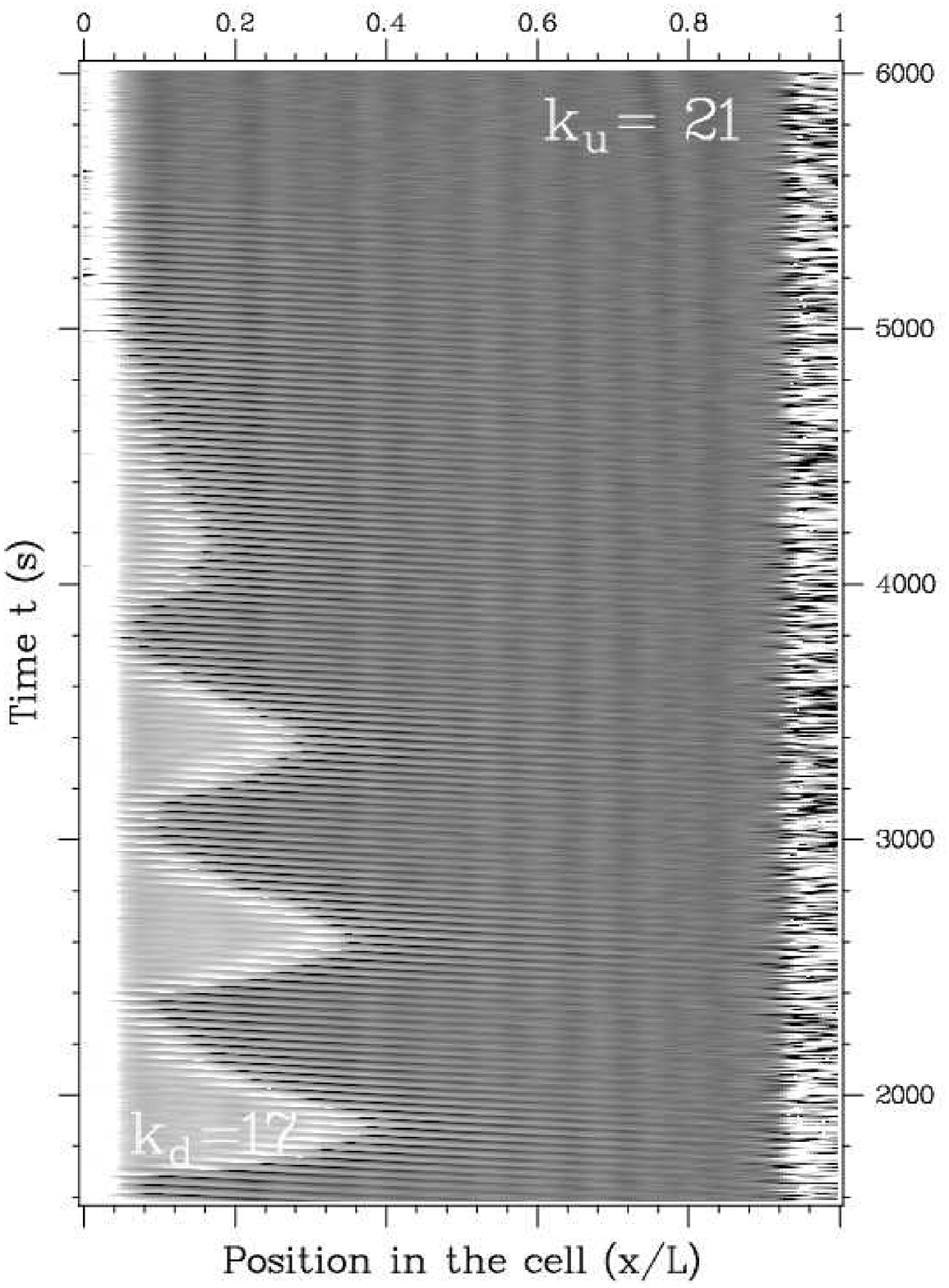}
\end{center}
\caption{Spatio-temporal diagrams of the spatial phase gradient (local 
        and instantaneous wavenumber) for transients leading to an 
        homogeneous state as the one of Fig.~\ref{fig:conv-abs}a.
        Initial conditions follow the protocol described in the text,
        they have been prepared at $t=0$. A dislocation front is 
        slowly advected out of the cell. The modulations grow along $x$
        but vanish along $t$: this is the signature of a convective 
        instability regime.
        Left: right-traveling wave for$\epsilon=0.79$ ($\Delta T=5.44$K), 
	the front position moves monotonically and leads to the final stage 
	presented in Fig.~\ref{fig:conv-abs}a. 
	Right: left-traveling wave $\epsilon=0.78$ ($\Delta T=5.52$K), the front position
        oscillates before being evacuated at a roughly constant speed.}
\label{fig:conv} 
\end{figure}

Figs~\ref{fig:conv-abs} and \ref{fig:conv} presents the three states
which support our discussion. For $\epsilon > \epsilon_{\rm m,a} = 0.79$
or $\Delta T > \Delta T_{\rm m,a} = (5.56 \pm 0.03)$K, the observed
pattern (Fig.~\ref{fig:conv-abs}b) can be described as a wave composed
of two wave-trains of mean wavenumbers $k_{\rm u}$ and $k_{\rm d}$. The
wavenumber, frequency and amplitude of both wave-trains are modulated in
space and time. The wavenumber $K_{\rm mod}$ of the modulation is of
order of $|k_{\rm u}-k_{\rm d}|$. Waves are emitted by one side of the
cell with wavenumber $k_{\rm u} \sim 21 \! \cdot \! (2\pi/L) \simeq 0.73$
mm$^{-1}$ and propagate along the cell at the phase velocity. The phase
modulation of this wave-train, traveling at the group velocity, is
spatially growing. in Fig.~\ref{fig:conv-abs}c, we clearly see the
exponential growth of the local-wavenumber modulation amplitude $A_{\rm
mod}$ along $x$. At a fixed, finite distance $x_F$ from the
source-boundary, the wavenumber modulation is so large that it allows
the wavenumber to change from $k_{\rm u}$ to $k_{\rm d}$ by
time-periodic phase slips. For $x > x_F$, the mean wavenumber is $k_{\rm
d} \sim 17 \! \cdot \! (2\pi/L) \simeq 0.59$mm$^{-1}$. In this second
region, the modulation is damped (Fig.~\ref{fig:conv-abs}c): we conclude
that $k_{\rm u}$ (resp. $k_{\rm d}$) waves are unstable (resp. stable)
with respect to modulations.

\begin{figure}
\begin{center} \includegraphics[width=8cm]{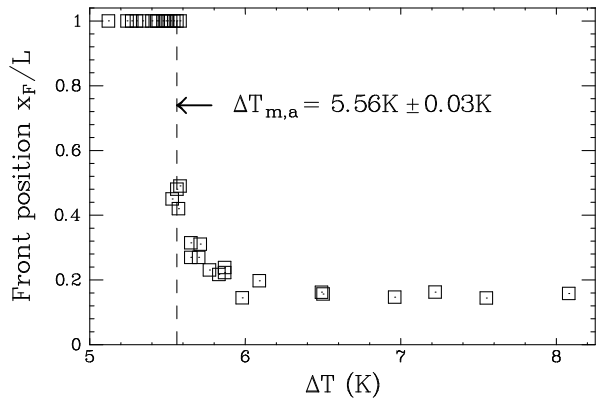} \end{center}

\caption{Spatial position $x_F$ of the dislocation front for absolutely
        unstable states {\it vs} $\Delta T$. Stable and convectively unstable
        states without permanent dislocation front are represented as
        realisations at $x_F=L$ (arbitrary choice).}

\label{fig:front}
\end{figure}

We call {\em dislocation front} the set of spatio-temporal locii where
spatio-temporal dislocations occur. For $\epsilon > \epsilon_{\rm m,a}$,
the position $x_F$ of this object is stationary; Fig.~\ref{fig:front}
shows the relation between the control parameter and the front position
which remains located in the first half of the cell whatever $\epsilon$.
Steady dislocation fronts have already been observed for traveling waves
in a Taylor-Dean experiment \cite{botmut00}. In general, hysteresis as
not been investigated. From the modulation amplitude profiles $A_{\rm
mod}(x)$ (Fig.~\ref{fig:conv-abs}c), we also extract the spatial growth
rate of the modulations: this quantity will be discussed below togheter
with convective and stable regimes (Section~\ref{sec:rec:conv_stab} and
Fig.~\ref{fig:ksi}).

We pretend those stationary states to be the global modes for the
modulational Eckhaus instability. All perturbations leave the
structure of those modes unaffected; the front position is always the same at a given
value of $\epsilon$. The structure of these global modes is very
original: nothing seems to saturate the modulations except the break-up of
the underlying wave-pattern, i.e., the abrupt change of the
mean-wavenumber downstream the dislocations. Similar patterns have been
observed numerically in semi-infinite \cite{Couairon:99} and closed
cells \cite{tobpro98}. Like Couairon and Chomaz \cite{Couairon:99} we
observe the nonlinear global threshold and the absolute instability
threshold to be identical.

\subsection{Convective instability states}

For $\Delta T < \Delta T_{\rm m,a}$, dislocation fronts are not observed
on asymptotic states. The asymptotic regime (Fig.~\ref{fig:conv-abs}a)
is an homogeneous wave, of uniform unmodulated wavenumber $k_{\rm u}
\sim 21 \! \cdot \! (2\pi/L)$. However, transients obtained after
control parameter changes show propagating dislocation fronts
(Fig.~\ref{fig:conv}). These fronts are slowly advected out of the
channel: those states are convectively unstable states with respect to
the modulational Eckhaus instability. Note that the front position can
increase monotonically or with an oscillation at a given low frequency.
It seems that this oscillation may result from the bouncing of
the front on the boundary. Although, there is no possible reflection
of the modulation on the boundary because there is not support
for a backward traveling modulation on the right-traveling carrier.
In
both cases, we measured the velocity of the front at the end of the
process where the velocity is almost constant. Those moving objects
are observed in the small gap between
$\epsilon_{\rm m,c}=0.76$ ($\Delta T_{\rm m,c} = 5.45$K) and
$\epsilon_{\rm m,a}=0.79$ ($\Delta T_{\rm m,a} = 5.56$K).

\subsubsection{Noisy source states}
\label{sec:rec:noisy}

Due to the special hydrodynamic of the thermocapillary problem, there is
another way of observing transients and measuring spatial growth rates.
The silicon oil used in the experiments is very volatile and during long
experimental runs, the fluid depth in the cell decreases. Of course, we
keep working with fluid depth almost constant around $h=1.7$mm, but
during some runs $h$ can decrease down to 1.65~mm. For this
small depth --- small but still within the error-bars we allowed --- the
boundary emitting waves, acting as a source, becomes larger in space and
noisy. This may be interpreted as follows: the system has at the
beginning of the run a rather well-fixed wavenumber --- around
21.($2\pi/L$) --- and wants to keep locked on this value for a given
$\epsilon$, even if the fluid depth $h$ decreases; this is only possible
if the effective size of the system is reduced by shifting the source
from the boundary of the cell to the interior of the cell. This gives an
apparently large source because the opposite traveling wave is very
damped and almost invisible. The position of the displaced source is not
well defined and this may explain the fluctuations observed. Moreover,
the source may emit modulations, so we call it a noisy source.

\begin{figure}
\begin{center}  \includegraphics[width=6.5cm]{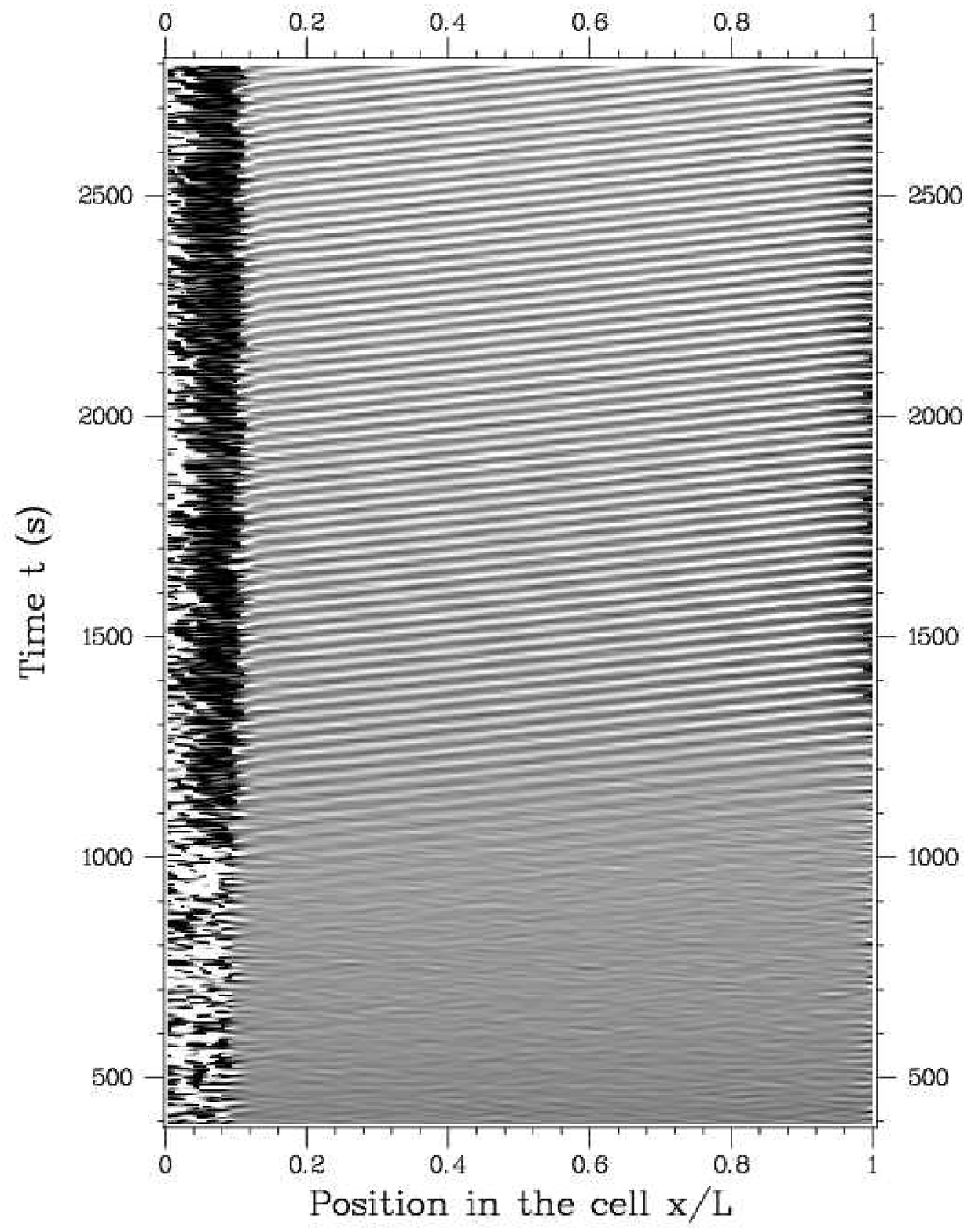}
                \includegraphics[width=6.5cm]{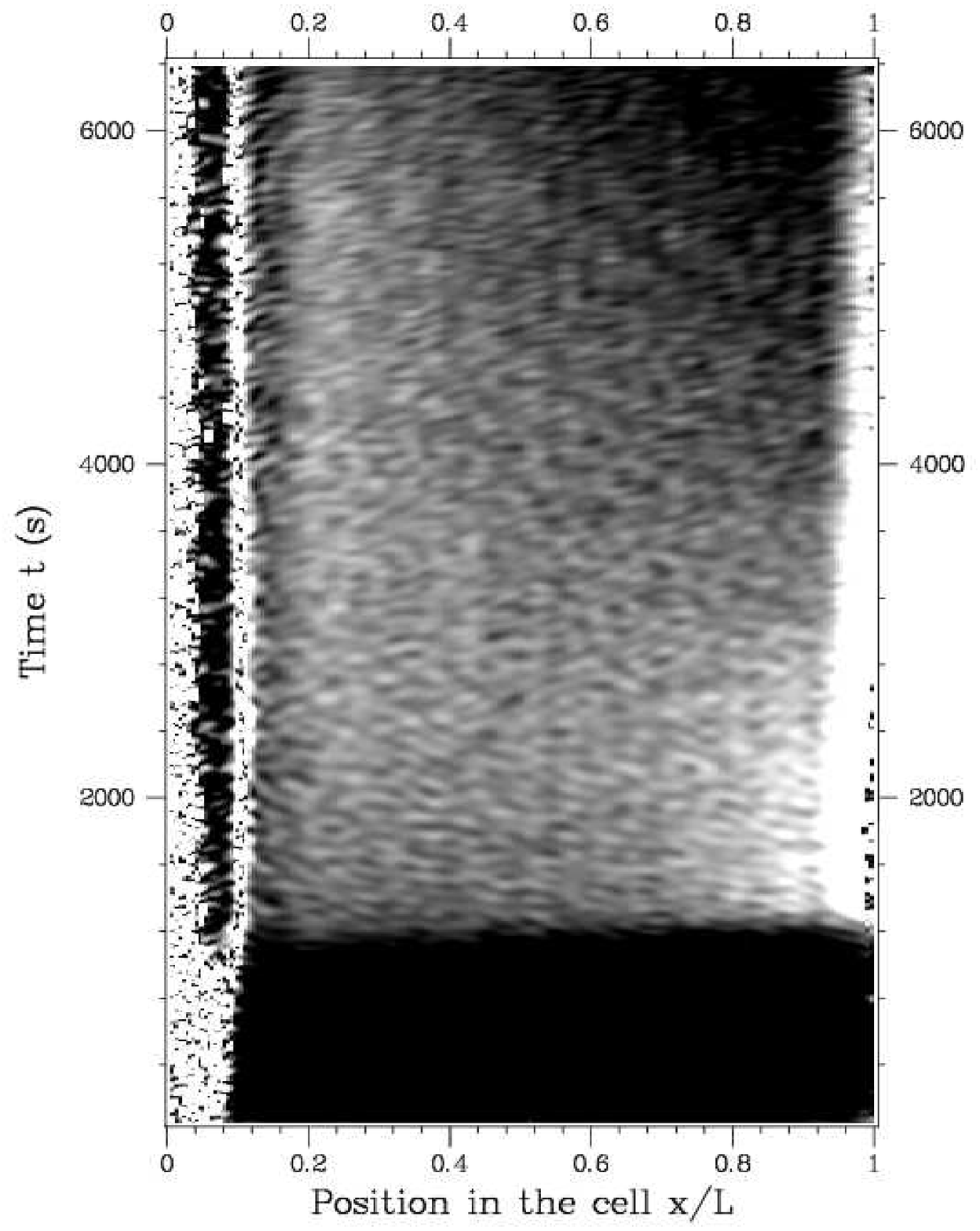}
\end{center}

\caption{Modulations emitted by a noisy source for $\epsilon=0.70$
($\Delta T=5.27$K). The noisy source appears at time $t\simeq 1000$s at
$x/L \simeq 0.1$. Left: spatio-temporal diagram of the local frequency
obtained after using Hilbert demodulation ($400{\rm s}< t < 2800{\rm
s}$). Right: spatio-temporal diagram obtained after a second
demodulation applied on the left-side diagram, showing the amplitude of
the modulation ($0 < t < 6400$s). Before time $t\simeq 1000$s,
the right-traveling hydrothermal wave is unmodulated: uniform
gray-level for the frequency (left) and black or zero amplitude for the
modulation (right). After $t\simeq 1000$s, the wave is modulated. One
can see on the amplitude of the modulation (right diagram) that, after a
transient ($t \gtrsim 4000$s), the perturbations emitted by the source
are spatially and temporally damped: this is illustrated by the darker
zone in the upper right part of the image. It is the signature of the
hydrothermal wave being stable with respect to modulations.}

\label{fig:st_noisy}
\end{figure}

Although hydrodynamic considerations may better explain the exact nature
of this source, we only look at it in the present study as a
source of modulations over the upstream wavenumber $k_u$, the stability of which
we are considering. The aim is to study the response of our system under
a continuous forcing and such a source provides a convenient
forcing. Fig.~\ref{fig:st_noisy} shows the response of the system to the
apparition of a noise source. 
On Fig.~\ref{fig:st_noisy}, a uniform right-traveling wave is displayed
for $t \lesssim 1000$s. At $t \simeq 1000$s, $h$ decreases
under a critical value and a source appears near $x/L \simeq 0.1$.
Then, modulations are continuously emitted from the source. On
Fig.~\ref{fig:st_noisy} (left) these modulations appear as waves on the
local frequency data. The amplitude of this modulation is viewed on
Fig.~\ref{fig:st_noisy} (right) obtained after demodulation of the left
diagram: black represents unmodulated waves (zero amplitude) and white
represents the maximum modulation level reached. After a complex
transient between $t \simeq 1000$s and $t \simeq 4000$s, we observe the
modulation amplitude to decay along both spatial and temporal axis.
In this case, the basic unmodulated waves are thus stable with respect
to the modulations induced by the presence of the noisy source.

\begin{figure}
\begin{center}
\includegraphics[width=8cm]{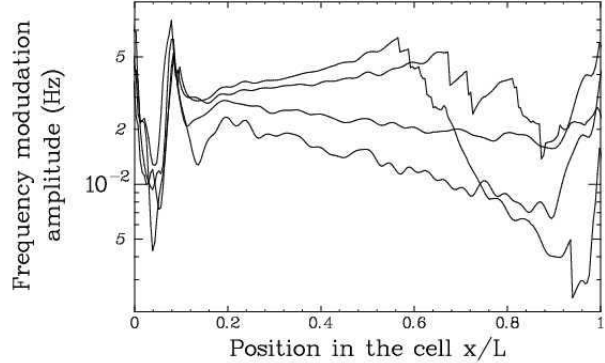}
\end{center}

\caption{Superposition of several asymptotic profiles of modulation
        amplitude in the case where a noisy source is present around. 
	The source is on the left ($x=0$) and the waves are right-propagating.
        Those profiles are extracted at the end of the time series, but still during
        the transient which may take a very long time to completely relax to zero.
	(see Fig.~\ref{fig:st_noisy} ,right).
	The signal portion just downstream of the source core ($x/L \gtrsim 0.2$)
	reveals the stability of the modulation.
        From bottom to top, $\epsilon=0.690$, $0.732$, $0.752$ and $0.765$.
        For the two lower values of $\epsilon$, the modulations relaxes to zero
        as they propagate: the unmodulated wave is stable.
	For the two higher values of $\epsilon$, the modulations are spatially 
	amplified: the unmodulated wave is convectively unstable.   
        The critical value 
        for the transition is $\epsilon_{\rm m,c}=0.758 \pm 0.010$
        or $\Delta T_{\rm m,c}=(5.45 \pm 0.03)$K.}

\label{fig:profiles:stable_convective}
\end{figure}

In Fig.~\ref{fig:profiles:stable_convective} several profiles of the
modulation amplitude are presented for various $\epsilon$. Such profiles
are extracted at the end of diagrams similar to Fig.~\ref{fig:st_noisy}
(right), {\em i.e.}, in the asymptotic state where the modulation
amplitude decays along time. Again, the amplitude of the frequency
modulation is obtained by a second Hilbert transform performed on the
local frequency data. We observed that for $\epsilon > \epsilon_{\rm
m,c}$ perturbations are spatially amplified whereas they are damped for
$\epsilon < \epsilon_{\rm m,c}$. 

\subsection{Stable / convectively unstable transition}
\label{sec:rec:conv_stab}

For $\epsilon < \epsilon_{\rm m,c}$, asymptotic states are uniform and
dislocation fronts do not exist in the absence of forcing. In the
presence of forcing by a noisy source, the perturbations are spatially
damped as they are advected away from the source
(Fig.~\ref{fig:st_noisy}). Close to $\epsilon_{\rm m,c}$ very long
transients are often observed. These transient patterns are also
slightly modulated; this is illustrated in Fig.~\ref{fig:st_noisy}. Most
often, the modulations do not reach the critical amplitude producing
dislocations. Asymptotically, the modulation amplitude is decreasing
(negative spatial growth rate) along the downstream direction for
$\epsilon < \epsilon_{\rm m,c}$, and increasing for $\epsilon >
\epsilon_{\rm m,c}$. Anyway, since $\epsilon < \epsilon_{\rm m,a}$, we
observe the modulation amplitude to decay along time. On the modulation
amplitude image in Fig.~\ref{fig:st_noisy} (right), a dark corner on the
upper right part on the image signs the spatial and temporal damping of
the modulation after a long relaxation. The complete relaxation may last
much longer than the experimental running time, and those results have
to be considered with care. Above $\epsilon_{\rm m,c}$, asymptotic states
in the presence of forcing are not uniform but constitute global modes
under an external forcing.

\begin{figure}
\begin{center}
\includegraphics[width=12cm]{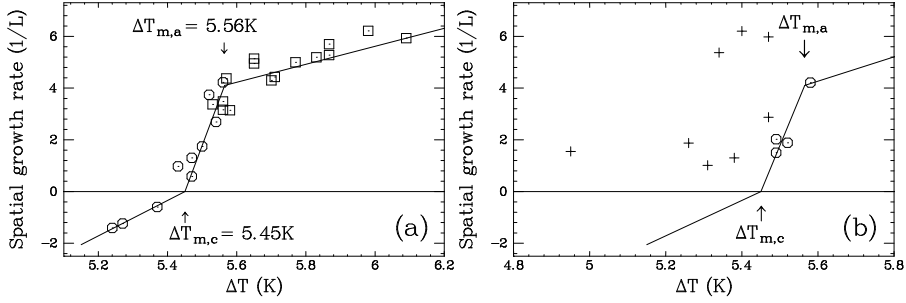}
\end{center}

\caption{(a) Evolution of the spatial growth rate of the modulation with
        the control parameter for spontaneously modulated wave patterns,
        transient ({\large $\circ$}) or steady ({\tiny $\Box$}). Linear fits of
        the three regimes ---stable, convective and absolute--- are presented.
        They intersect at $\epsilon_{\rm m,c}$ and $\epsilon_{\rm m,a}$. These
        data concerns the modulations of the upstream region of the cell whose
        mean wavenumber is $k_{\rm u} \sim 21 \! \cdot \! (2\pi/L)$.
        Corresponding data for the downstream region are negative while $\Delta
        T \le 8$K. (b) Idem for perturbation initiated wave-packets in the
        stable ({\footnotesize $+$}) and convectively unstable ({\large
        $\circ$}) regimes. The solid lines recall the fits of (a) to allow
        quantitative comparisons: the same selection of the growth rate is
        observed in both cases for the convective regime.}

\label{fig:ksi}
\end{figure}

On a quantitative point of view, we measured spatial and temporal growth
rates of the modulation. We will present these data for the unstable
upstream wave-train. The {\em temporal} growth rate for modulations in
the laboratory frame is negative below $\epsilon_{\rm m,a}$ and positive
above. It is also close to zero around the convective transition where
very long transient are reported. The spatial growth rate of the
upstream $k_{\rm u}$ wave-train for all three regimes is presented on
Fig.~\ref{fig:ksi}a. It is positive for both unstable states but the
slope is seemingly different in the convective and absolute regimes. It
is negative below $\epsilon_{\rm m,c}$.

\subsection{Perturbed states}

\begin{figure}
\begin{center}  \includegraphics[width=6.5cm]{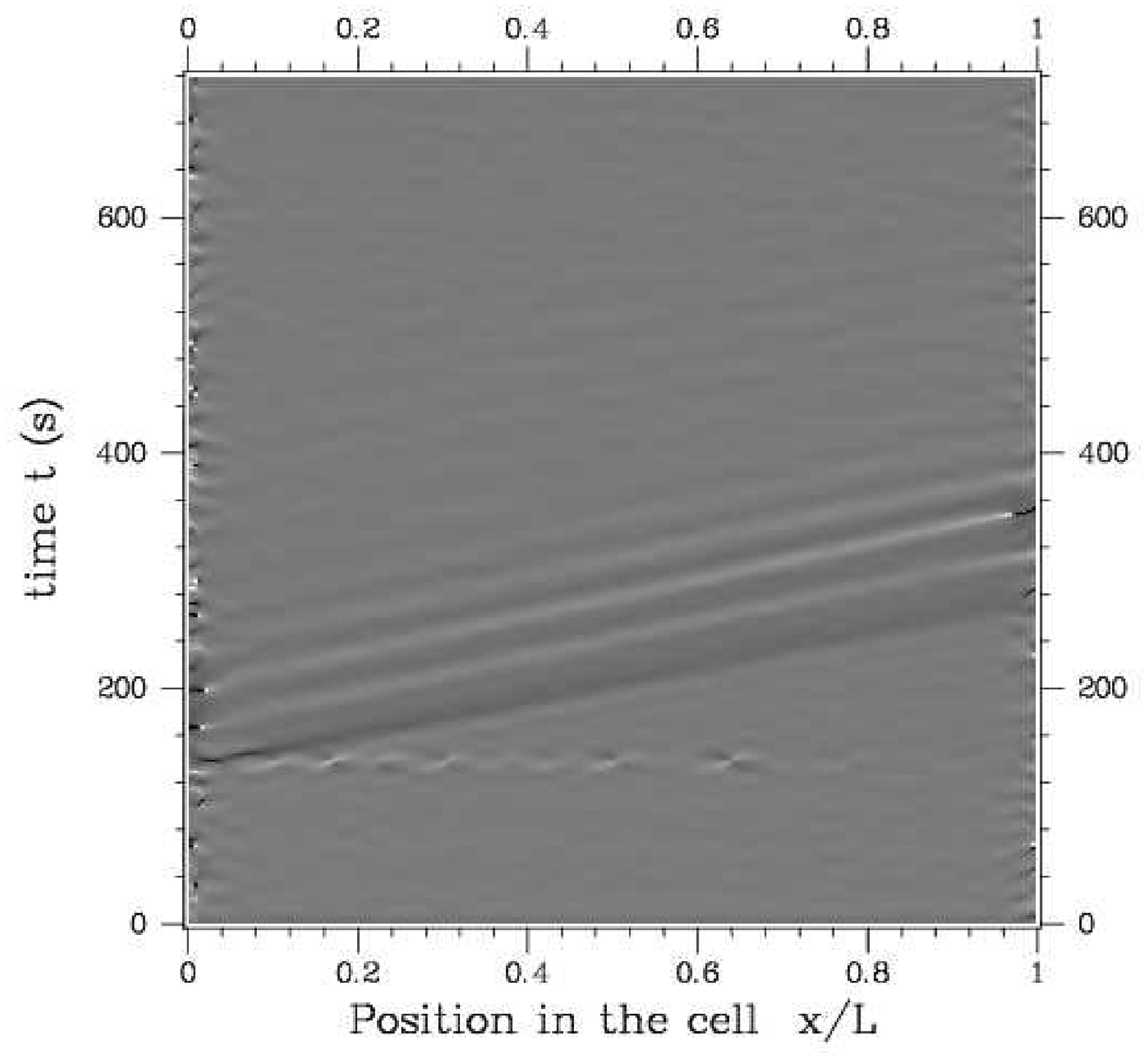} 
                \includegraphics[width=6.5cm]{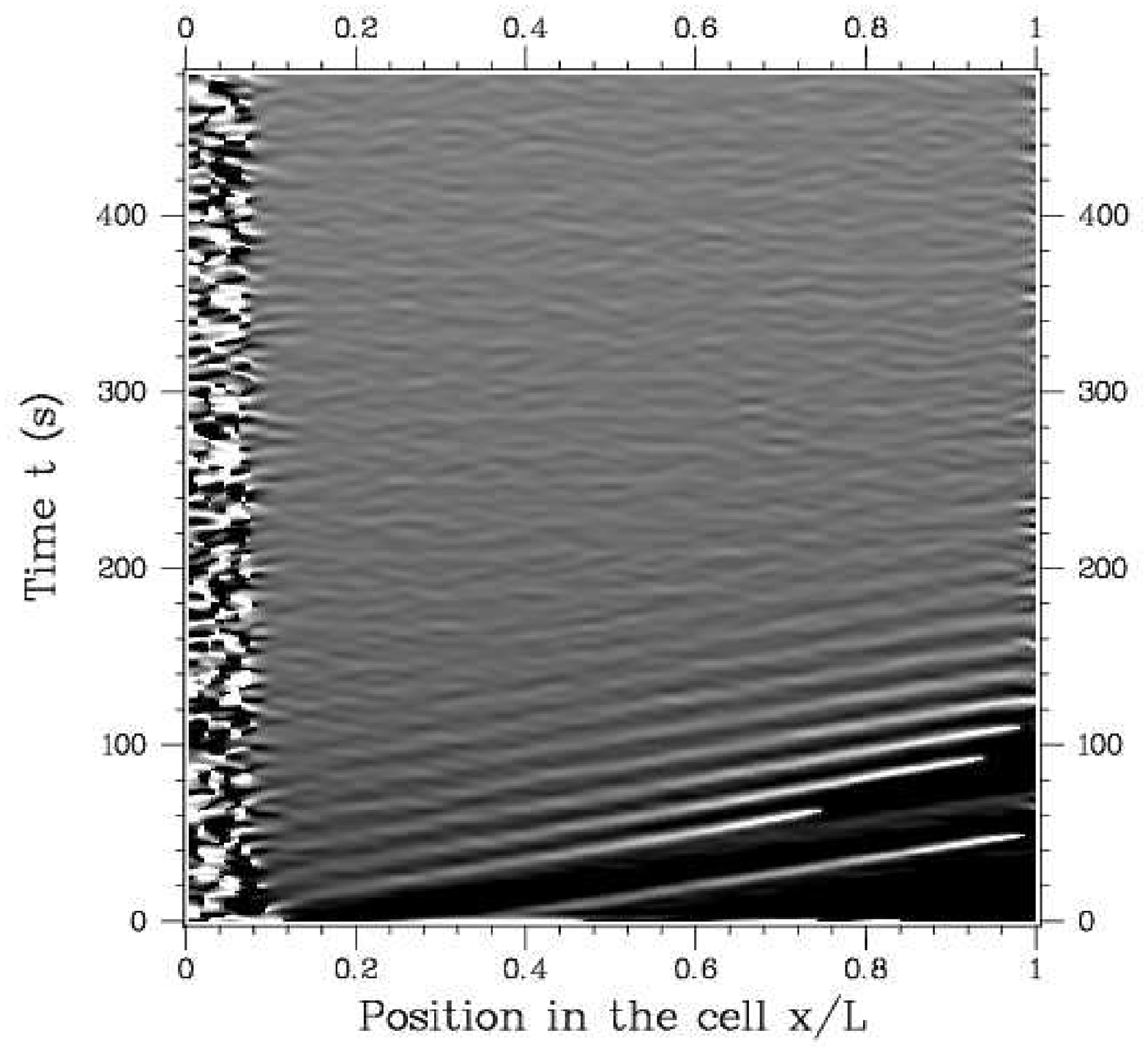} \end{center}
\caption{Spatio-temporal diagrams of the local frequency showing
        the response to an external mechanical perturbation.
        Left:  $\epsilon=0.735$ ($\Delta T$=5.38K); 
               a perturbation is induced at time $t=130$s. 
        Right: $\epsilon=0.764$ ($\Delta T$=5.47K);
               a perturbation is induced at time $t=0$s. 
        Modulations are spatially growing, but are advected away very quickly.
	So they disappear after a few 100 seconds. }

\label{fig:st_perturbated}
\end{figure}

In order to test the above description, we perturbed the uniform states
by either plunging a thin needle in the convective layer or by dropping
a cold or hot droplet of fluid. Two examples are reproduced on
Fig.~\ref{fig:st_perturbated}. The frequency contents of those
perturbations differ from the above reported transients or forced
states: the modulation wave-trains contains only a few wavelengths and
appears to be advected downstream at roughly the group velocity. All
observed perturbations show positive spatial growth rate and negative
temporal growth rate in the laboratory frame. The spatial growth rates
are presented in Fig.~\ref{fig:ksi}b. In the convective regime, the
growth rate appears to be selected at the same value than in spontaneous
transients or forced states. In the stable regime, however, the data are
very dispersed but remain positive.

\subsection{Discussion}

\begin{figure}
\begin{center} \includegraphics[width=8cm]{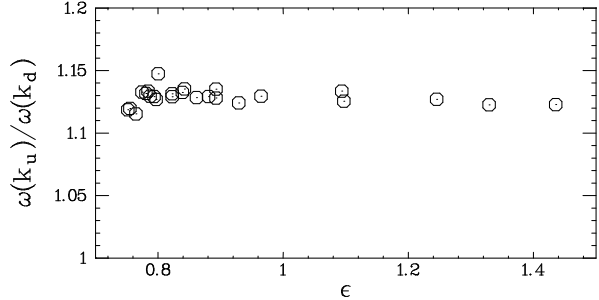} \end{center}
\caption{Ratio of the frequency in the upstream region by the frequency
        in the downstream region, {\em versus} $\epsilon$.
        This ratio is almost constant, and its value $1.130 \pm 6.10^{-3}$
        is
far from any simple fraction.}
\label{fig:incommensurate}
\end{figure}

First, let's point out that the upstream and downstream wavenumber are
incommensurate, and so are the associated frequencies. On
Fig.~\ref{rec_f-k}, two branches are visible on the frequency data,
corresponding to the two different wavenumbers $k_{\rm u}$ and $k_{\rm
d}$. The mean ratio between the frequency of $k_u$ and $k_d$ is $1.130
\pm 6.10^{-3}$, {\em i.e.}, far from any simple rational fraction. This is
shown in Fig.~\ref{fig:incommensurate}. We then conclude than those two
frequencies are not resonant and that no locking occurs in our system.

Second, let's point out that the modulation amplitude $A_{\rm mod}$
never saturates. All observed $A_{\rm mod}$ profiles appear locally
exponential along $x$. No non-linear saturation effect is thus observed.
The dislocation onset, for a given $A_{\rm mod} \sim |k_{\rm u}-k_{\rm
d}|$ is the only limit to exponential growth. This is a strong argument
for the Eckhaus instability to behave subcritically in this closed cell.
Remember it is supercritical in the annular cell \cite{mukchi98} (see I
for discussion).

Third, the modulation wave system is a perfect single wave system:
reflection of the modulations at the boundaries are irrelevant for there
is no possibility for reflected information to travel back to the source
without being supported by a counter propagating wave.

The observation facts related above are coherent with the interpretation
in terms of convective and absolute instability. The striking point is
the positive spatial growth rates for perturbations in the seemingly
stable regime below $\epsilon_{\rm m,c}$. As for spontaneously modulated
patterns, we would expect those modulation wave-packets to decrease in
space exactly as the stable $k_{\rm d}$ wave-trains do in the absolute
regime (Fig.~\ref{fig:conv-abs}c).

\begin{figure}
\begin{center} \includegraphics[clip,width=8cm]{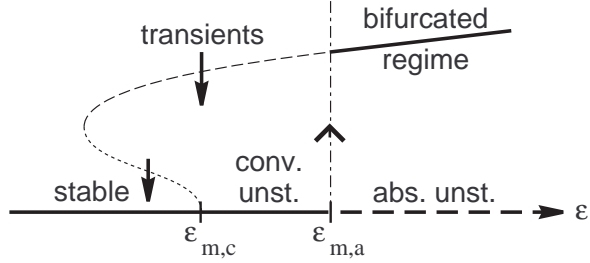} \end{center}
\caption{Schematic representation of the observed regimes, based on the
        usual representation of a subcritical bifurcation with zero group
        velocity. The control parameter $\epsilon$ is in abscissa, while the
        ordinate is only qualitative. Solid heavy lines represents the steady
        states, bifurcated or not, above or below $\epsilon_{\rm m,a}=0.794$K.
        The thin dashed lines may account for two different transient modes
        (see text).}
\label{fig:map}
\end{figure}

We may explain this in the following way: Suppose the convective
instability is subcritical as suggested previously. Then, above
$\epsilon_{\rm m,c}$, the transient evolves on an unstable branch
(Fig.~\ref{fig:conv}), close to the absolute branch
(Fig.~\ref{fig:conv-abs}b). However, below $\epsilon_{\rm m,c}$, a
second unstable branch co-exists, which can be reached only by
perturbing the flow: this description can be supported by the schematic
Fig.~\ref{fig:map} inspired by zero group velocity instabilities. These
branches present very different patterns: The upper branch exhibits
extended modulations over the whole cell, with slow evolution and, for
high enough amplitudes ---the generally observed case above
$\epsilon_{\rm m,c}$---, dislocation fronts. The lower branch exhibits
fast traveling narrow modulation wave-trains and cannot be reached
spontaneously by varying $\epsilon$.

This hypothesis can explain the very different aspect of spontaneous and
induced transients in the stable regime below $\epsilon_{\rm m,c}$. It
is also known that the shape of induced non-linear patterns below
subcritical instabilities depends of the forcing amplitude
\cite{botcha98}, so the dispersion in Fig.~\ref{fig:ksi}b may be due to
both the effect of amplitude and the presence of two branches.

\begin{figure}
\begin{center} \includegraphics[width=8cm]{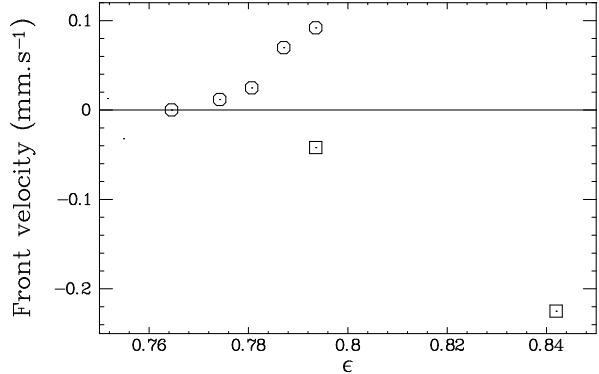} \end{center}
\caption{Front velocity around the convective/absolute transition. The
        circles ({\large $\circ$}) show the velocity of dislocation fronts in
        transient convective regimes below $\epsilon_{\rm m,a}$. The (negative)
        velocities of transient modulation fronts invading the cell from
        downstream, above $\epsilon_{\rm m,a}$, are shown by squares ({\tiny
        $\Box$}). For comparison, the group velocity is $0.90 {\rm mm.s^{-1}}$.}
\label{fig:vfront}
\end{figure}

Another observation of the convective branch is intriguing. We report in
Fig.~\ref{fig:vfront} the asymptotic velocity of the dislocation fronts
between $\epsilon_{\rm m,c}$ and $\epsilon_{\rm m,a}$, i.e., the tangent
to the space-time trajectory when the front quits the cell as on
Fig.~\ref{fig:conv}. The observation is surprising: the closer the
absolute instability onset, the faster the front moves! And then jumps
below zero above $\Delta T_{\rm m,a}$. {\em A contrario}, around
$\epsilon_{\rm m,c}$, the front velocity is zero, leading to infinitely
long transients, i.e., temporal marginality. This quantifies the
experimental complexity of carrying the experiment around this point.

What is the meaning of the velocity jump at $\epsilon_{\rm m,a}$? Is the
convective/absolute transition also subcritical? Probably it is: while
our protocol did not allow to explore all branches by varying $\epsilon$
up and down from one state to another, a test have been made to transit
directly from an absolute state to a stable state just below
$\epsilon_{\rm m,c}$: the absolute modulation profile remains fixed in
the cell. This can be due either to hysteresis, either to the vanishing
front velocity... which makes the system marginal in this region. This
point would need to be addressed with an improved experimental device.

\begin{figure}
\begin{center} \includegraphics[width=8cm]{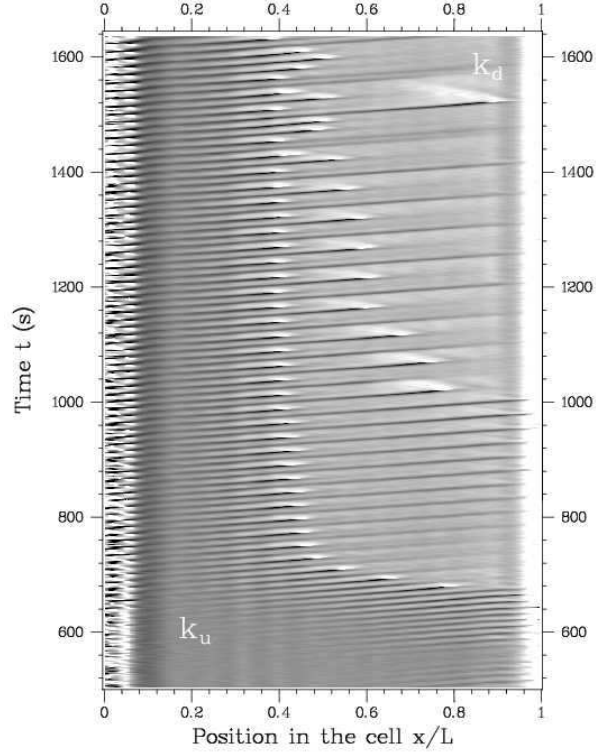} \end{center}
\caption{Transient for $\epsilon=1.74$ with $k_u=19\cdot(2\pi/L)=0.65$mm$^{-1}$
        and $k_d=16\cdot(2\pi/L)=0.56$mm$^{-1}$.
        A period doubling occurs for the modulation front: one modulation
        over two explodes in a dislocation forming a first dislocation
        front in space-time, then one modulation over 4 explodes
        forming another dislocations front, then one over 8 forms another front,
        and so on until all those dislocations fronts merge together 
        leaving a state in which $k_u$ is absolutely unstable.}
\label{fig:doubling}
\end{figure}

Let's note that for higher $\epsilon$ values, the front position
$x_F(t)$ exhibits chaotic behaviors and can thus be viewed as the order
parameter for the modulational instability up to the transition to
spatio-temporal chaos. An example of disordered state is presented on
Fig.~\ref{fig:doubling} which suggest that the spatio-temporal behavior
can be described using this front as a dynamical system.

Last but not least, modulations in the rectangle can be connected to
modulations in the annulus described in section~\ref{sec:annulus} of I.
First, in periodic boundary conditions, the Eckhaus secondary
instability is supercritical for wavenumbers close to the critical one,
whereas it is rather subcritical far from the critical wavenumber
$k_{\rm c}$. This is confirmed by observations in non-periodical
boundary conditions: the selected wavenumbers are far from $k_{\rm c}$
in the rectangle and the modulational instability is subcritical. This
opens the following question: can the observed modulations in the
rectangle be described as modulated amplitude waves
(MAWs)~\cite{brutor01a,brutor01b,hecke01} ? Such modulations have
already been seen by~\cite{kol92,baxeat92,liueck99}. As pointed out
before, modulated waves in the rectangle are not saturated, neither
should be corresponding MAWs; moreover, the spectral richness of the
modulations is weak in the rectangle: modulations are almost
monochromatic. So, speaking of MAWs, we are facing solutions of the CGL
equation that connect two non-saturated modulated waves.

\section{Sources and sinks in the annulus}
\label{sec:sources}

Coming back to the annulus will allow us to show that the
convective/absolute transitions advocated for in the rectangle is also
relevant in periodic boundary conditions, when the Galilean invariance
is broken due to the presence of both right- and left-traveling waves.
The interpretation of bifurcations in the rectangle in terms of
convective/absolute transitions is then reinforced.

\subsection{Obtaining source/sink pairs}

Let us describe how patterns form when $\epsilon$ is rapidly increased from a
negative value to a supercritical value $\epsilon_{\rm f}$; describing
those transients allow us to distinguish between different behaviors.
In all such experiments, the waves first appear in small patches at
several places in the cell. The envelopes of those patches propagate at
group velocity while their spatial extension increase. After a short
transient, waves have invaded all the cell but sources and sinks are
presents which are reminiscent of the boundaries of the initial patches.
The number of initial patches, and so the number of sources and sinks
depends on the time derivative of the ramp in $\epsilon$ leading to
$\epsilon_{\rm f}$ : the faster the control parameter is varied across
the threshold of waves, the more sources/sinks pairs are present in the
cell; a typical realization shows two or three pairs. Quasi-static
variations of $\epsilon$ shows at least one pair.

\begin{figure}
\begin{center}  \includegraphics[width=6.5cm]{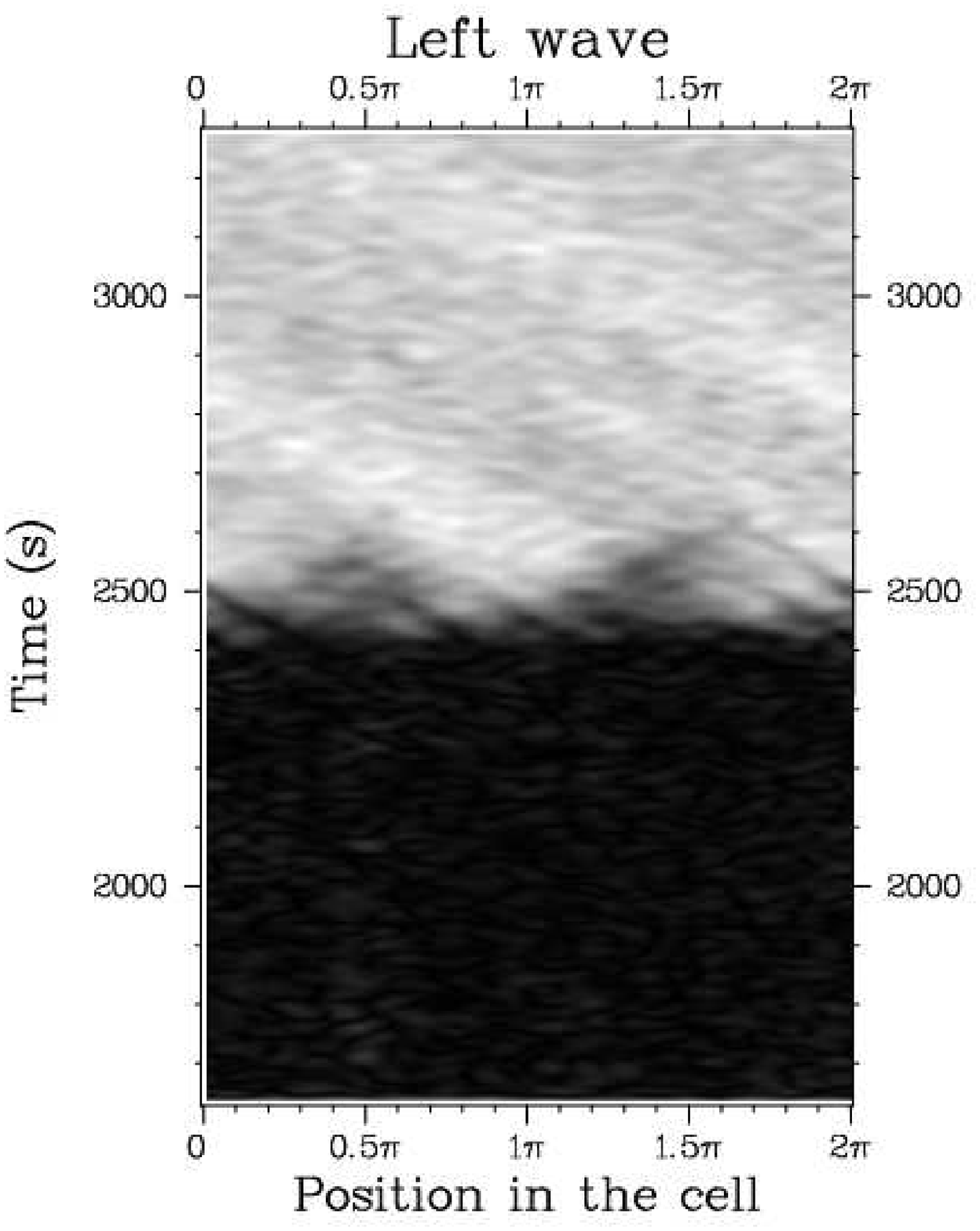}
                \includegraphics[width=6.5cm]{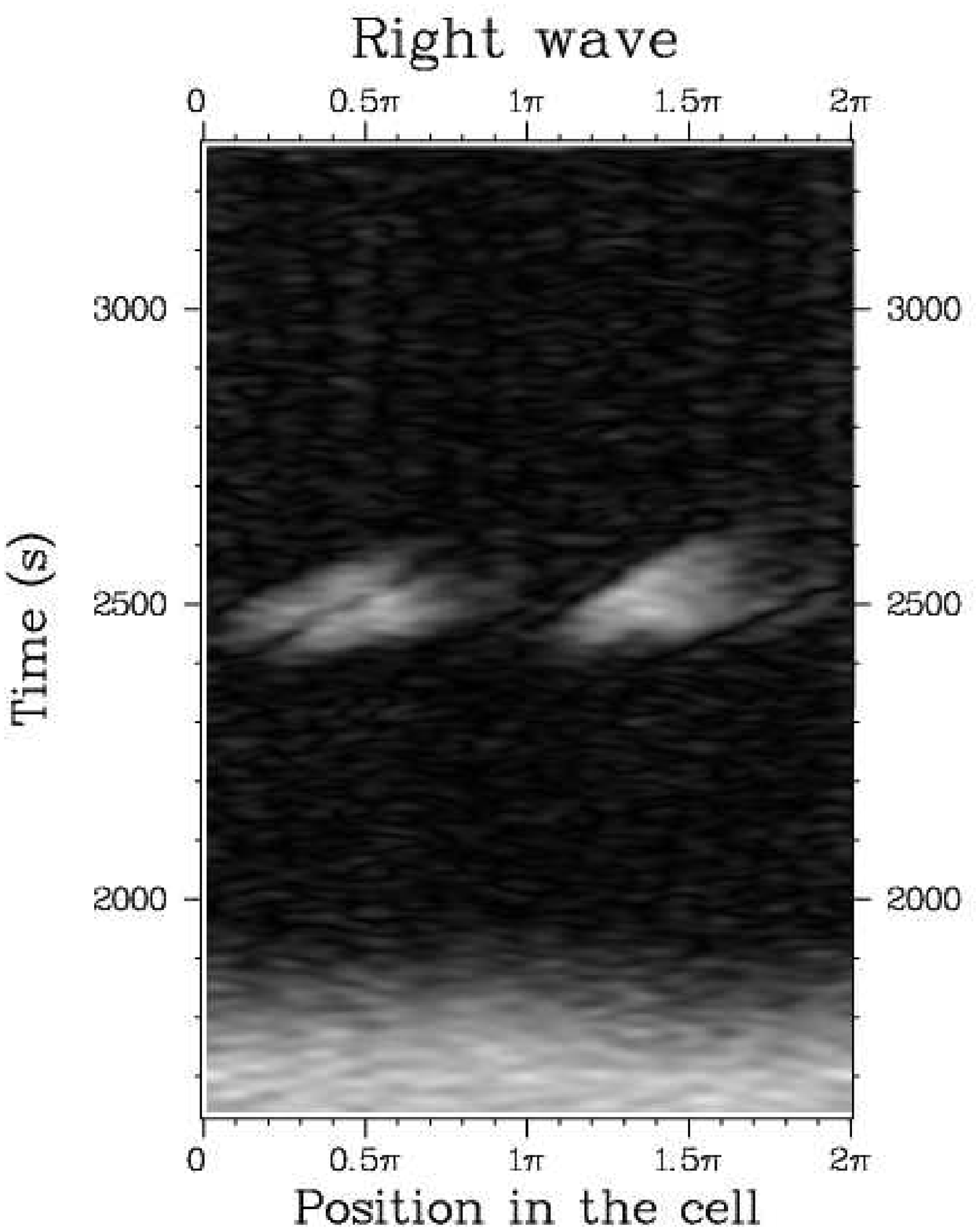}
\end{center}

\caption{Spatio-temporal diagrams of the local amplitude showing
        the initial competition between right- and left-propagating waves
        in the annulus.
        Left : amplitude of left-traveling waves.
        Right : amplitude of right-traveling waves.
        The gray scale is proportional to the amplitude (increasing from black
	to white).
        $\Delta T$ is increased slowly from a (slightly) subcritical value
        to 3.60 K ($\epsilon_{\rm f} = 0.16$). 
	Two couples of source/sink appear quickly at time $t=2400$s;
        then they interact and only a single wave remains after time $t=2600$s.
        Note that initially (before $t=1800$s), only the right-traveling wave
        is present, and after $t=2700$s, only the left-traveling wave is present;
        on that remaining left wave, a transient modulation is present.}

\label{fig:st_competition}
\end{figure}

This first "invasion" time is followed by a second transient regime
where sources and sinks interact. Fig.~\ref{fig:st_competition} presents
a typical competition leading to the annihilation of two source/sink
pairs. For this experiment, we start from a supercritical value
$\epsilon = 0.14$ ($\Delta T=3.52$K) and a single right-traveling wave
and then reduce the control parameter by switching off the electrical
heating of the inner block. So $\Delta T$ decreases and the
right-traveling wave disappear as the onset is crossed from above. As
soon as the wave disappear, we switch the heating on again. Waves
reappears in four different patches forming two sources and two sinks.
Then, after a 5 minutes transient, the left wave remains alone.

A nice observation is done by looking around the value $\epsilon_{\rm
a}$. If the operating value $\epsilon_{\rm f}$ is close to 0, all pairs
annihilate within the transient and the asymptotic pattern is always a
single homogeneous traveling wave. For $0< \epsilon_{\rm f} < \epsilon_{\rm
a}=0.18$, no source/sink pattern has been seen asymptotically. In
contrast, when $\epsilon_{\rm f}>\epsilon_{\rm a}$, {\em i.e.}, the ending
value of the imposed control parameter is highly supercritical, it is
possible ---but not mandatory--- to have a source/sink pattern frozen
on long times. We have observed that the higher $\epsilon_{\rm f}$ is,
the easier source/sink states are easily frozen~\cite{mukchi98}. For
$\epsilon_{\rm f} > \epsilon_{\rm a}$, we have observed some realizations in
which sources and sinks are present after times much longer than the
diffusion time.

So in the annular experiment, sources and sinks have been observed
during transients for all possible values of $\epsilon$. But a striking
result is that the couples of sources and sinks are always unstable
below $\epsilon < \epsilon_{\rm a}=0.18$; in that case, all source/sink
pair collapse and only an homogeneous single wave subsists. In contrast,
sources and sinks have been observed in a stable way for
$\epsilon > \epsilon_{\rm a}$, {\em i.e.}, at least such a pair lasts as
long as the experiment is performed.

\subsection{Convective/absolute onset in the annulus}

Below $\epsilon_{\rm a}$, source/sink couples are unstable in the
sense that they collapse together, but sources can also be qualified 
as unstable because they emit modulations. Those modulations are
also present close to the onset of waves.

It has been theoretically and numerically~\cite{hecsto99,ebesaa00}
established that stables sour\-ces only exists in a 1D-periodical
geometry when the control parameter $\epsilon$ is above $\epsilon_{\rm
abs}$, {\em i.e.}, when the instability is absolute. This may be
understood remembering that in the convective regime, the wave-system
acts as a noise amplifier. The two different regimes for the sources 
are to be discerned: below $\epsilon_{\rm abs}$, sources are noisy and
their width is large; above $\epsilon_{\rm abs}$, sources selection 
a wavenumber and have a smaller spatial extend. 

Measurements of the source width have been performed recently in a very
long rectangular geometry~\cite{paswes01,note:Pastur} leading to a nice
confirmation of this transition. Here, we use the qualitative
observation of stable sources above $\epsilon_{\rm a}=0.18$ and the
disappearance of all sources below $\epsilon_{\rm a}$ to confirm that
$\epsilon_{\rm abs}=0.18$ in the annulus. Using expression of
$\epsilon_{\rm abs}$ in Eq.~(\ref{eq:epsilon_abs}), and assuming that the
parameters involved are the same in both rectangle and annular geometry,
we are confident that the overall description using convective/absolute
transition for the onset shift in the rectangle is relevant.

\subsection{Discussion}

As previously said, modulations are emitted by the sources, even if
$\epsilon$ is close to 0. Moreover, when a single traveling wave is
finally produced by the collapse of the last sink within the last
source, modulations are also emitted. So the single unmodulated uniform
traveling wave pattern appears always after a transient in which initial
modulations are present but damped. This transient is studied in I
(section \ref{sec:mwdamped} and Fig.~\ref{fig:mwdamped}). The detailed
study of modulations emitted by the sources remains to be done, but one
can expect that sources in the annulus behave like the one in the
rectangle. For example, if their position is not well fixed as it is the
case of noisy sources in the rectangle (section~\ref{sec:rec:noisy}), it
may explain the same emission of modulations. 

As studied in Refs~\cite{hecsto99,riekra00} the effect of the group
velocity is also a key point for a better understanding of traveling
wave systems in both non-periodic and periodic geometries. Other
parameters such as the complex coupling coefficient ($\lambda+i\mu$ in
Eq.~\ref{eq:cgl:rec}) should also be tracked for influence on the
observed patterns in both geometries, thus suggesting that the complex
Ginzburg-Landau equations cannot be only described using simply two
parameters ($c_1$ and $c_2$ in Eq.~\ref{eq:cgl:rec}).

\section*{Conclusion}

Owing to their apparition via a supercritical instability with finite
frequency, finite wavenumber and finite group velocity, hydrothermal
waves were previously shown to be very well modelized by an amplitude
equation of the complex Ginzburg-Landau type (paper I). In the present
article, we used a one-dimensional hydrothermal wave system as an
experimental expression of the one dimensional CGL equation or of a
system of coupled one-dimensional CGL equations, to explore the effect
of the boundary conditions.

We have presented the global mode appearing in a rectangular box at the
absolute instability threshold for hydrothermal traveling waves.
Qualitative and quantitative comparisons have been performed to
distinguish from the case of a reflection-controlled global mode. The
relevance of the convective/absolute distinction was demonstrated by
accurate comparison of threshold values and critical behaviors of the
order parameters. Those measurements have revealed that the transition
is fully non-linear in the sense of Chomaz and Couairon \cite{chocou99}
and is well connected to the predictions of~\cite{tobpro98} for a single
wave pattern. 

For higher control parameter values in the rectangle bounded cell, we
observe a quasi one-directional traveling-pattern which undergoes an
Eckhaus secondary instability leading to traveling modulations. These
modulated patterns behave as non-linear fronts whose dynamics reveals
convective and absolute regimes as well \cite{garchi00,garchi02}. Thus
we have observed both stable/convective and convective/absolute
transitions for the modulational instability. The stable/convective
transition is subcritical and is characterized by zero spatial growth
rate for the modulation, together with zero advection velocity of the
modulated pattern, which can be viewed as spatial and temporal
marginality. The convective/absolute transition is characterized by the
dynamics of dislocation fronts. According to front velocity data, we
suggest this transition to be subcritical as well which qualitatively
differs from the case of the annulus. This question deserves a
theoretical support which remains yet, as far as we know, unexplored.

Finite group velocity in the presence of boundaries, leading to the
transition from convective to absolute, are linked with important
qualitative and quantitative changes of the global structure of the wave
patterns in rectangular geometry. We showed that this is also true in
the annulus where sources emitting waves are very sensitive to the
convective or absolute nature of the primary waves. The description of
one-dimensional hydrothermal waves using Ginzburg-Landau equations
appears to be very complete and satisfactory. One may expect even more
connections with future theoretical work on these equations.

\section*{Acknowledgments}

We wish to thank Lutz Brusch, Jean-Marc Flesselles, 
Joceline Lega, Carlos Martel, Wim van Saarloos, Luc Pastur,
Alessandro Torcini and Laurette Tuckerman for interesting discussions. 
Special thanks to Vincent Croquette for providing us his powerful
software {\sc XVin}. Thanks to Alexis Casner and Fr\'ed\'eric Joly
who contributed to the data acquisition, and to C\'ecile
Gasquet for her efficient and friendly technical assistance.

\bibliographystyle{unsrt} 
\bibliography{MWbiblio}	

\begin{thebibliography}{10}
\bibitem{note:Pastur}
In reference \cite{paswes01}, the geometry is non-periodical, but very
  extended. This allows to conjecture that $\epsilon_{\rm r} < \epsilon_{\rm
  abs}$ and so that waves exist for $\epsilon<\epsilon_{\rm abs}$.
\end{thebibliography}

\end{document}